\documentclass[preprint, authoryear]{elsarticle}
\journal{Advances in Space Research}

\date{}

\usepackage{lineno,hyperref,eurosym}
\modulolinenumbers[5]
\usepackage[dvipsnames]{xcolor}
\usepackage{moresize}

\usepackage{amsmath, mathtools, amssymb, bm}
\usepackage{algpseudocode}
\usepackage{siunitx} 

\usepackage{}

\usepackage{comment}

\usepackage{graphicx}
\usepackage{booktabs}
\usepackage{tabu}
\graphicspath{{images/}}
\usepackage{caption}
\usepackage{subcaption}
\usepackage{url}
\usepackage{setspace}
\usepackage{fancyhdr}
\usepackage{hyperref}
\usepackage{tabularx}
\usepackage{booktabs}
\usepackage{threeparttable}
\usepackage{url}
\usepackage{multirow} 
\usepackage{booktabs}
\usepackage{makecell}
\usepackage{natbib}
\let\cite\citep

\begin{document}

\begin{frontmatter}

\title{Conceptual Design and Analysis of a NanoTug Swarm for Active Debris Removal}

\author[label1]{F. Alnaqbi*}
\author[label1,label2]{S. Biktimirov}
\author[label3]{G. Gaias}

\address[label1]{Propulsion and Space Research Center, Technology Innovation Institute, Abu Dhabi (United Arab Emirates)}
\address[label2]{Department of Aerospace Engineering, Khalifa University of Science and Technology, Abu Dhabi (United Arab Emirates)}
\address[label3]{Department of Aerospace Science and Technology, Politecnico di Milano, Italy}

\cortext[]{Corresponding author: fatima\_alnaqbi98@hotmail.com (F.~Alnaqbi)}

\begin{abstract}
This paper investigates a swarm-based concept in which a number of nanosatellites, referred to as NanoTugs, are deployed by a mother spacecraft to capture and cooperatively stabilize and de-orbit space debris. The study focuses on the stabilization and de-orbiting phases of the mission, where each NanoTug is equipped with thrusters to perform the de-orbiting maneuver. An analytical method is developed to provide a preliminary understanding of the relationship between swarm system parameters, debris properties, and mission performance, which is subsequently verified through numerical simulations. Two NanoTug distribution strategies, random and predefined, are considered, and their influence on mission performance is evaluated. De-orbiting is achieved by thrusting along the direction that maximizes the reduction of the semi-major axis, as obtained from Gauss’ variational equations, while the attitude of the combined debris–NanoTugs system is controlled using a Lyapunov-based control law. A task allocation strategy is implemented to assign on/off commands to individual thrusters. Simulation results demonstrate the applicability of the analytical swarm sizing approach; however, a margin in system sizing is required due to the simplifying assumptions used in the first-order estimation. The proposed control approach for debris de-orbiting is shown to be feasible through representative mission simulations. In terms of NanoTug distribution across the debris surface, the predefined strategy provides improved performance, requiring fewer NanoTugs and offering more predictable behavior, whereas the random distribution results in frequent switching between NanoTug thrusters. Overall, the results highlight the feasibility of the swarm-based NanoTug concept for cooperative debris stabilization and de-orbiting.

\end{abstract}


\end{frontmatter}


\section{Introduction} 
\label{sec:introduction}
Active Debris Removal (ADR) in Earth orbit has become a critical area for the majority of the space agencies such as the European Space Agency (ESA), the National Aeronautics and Space Administration (NASA), the Japan Aerospace Exploration Agency (JAXA), and is also gaining attention from industry to support the long-term sustainability of space operations~\cite{NASA1999SpaceDebris, UNOOSA2025IADCguidelines}. The increasing amount of space debris poses a significant risk to operational satellites, raising the likelihood of in-orbit collisions and endangering future missions. In response, space agencies began introducing regulations and requirements mandating that newly launched satellites include end-of-life plans. These plans require either de-orbiting to very low altitudes or re-orbiting to graveyard orbits, away from protected orbital regions, after mission completion~\cite{ESA2025SpaceEnvironment}.

To support international collaboration on debris issues, the Inter-Agency Space Debris Coordination Committee (IADC) was officially established in 1993. The IADC facilitates the exchange of research and technology developments related to space debris mitigation ~\cite{UNOOSA2025IADCguidelines}. The IADC guidelines urge operators to limit debris formation during satellite operations, avoid on-orbit breakups or explosions, ensure spacecraft are removed from protected regions at end-of-life, and prevent collisions—especially within large constellations—to reduce the creation of new debris~\cite{UNOOSA2025IADCguidelines}.

A large number of debris currently in orbit are non-collaborative, meaning they lack the ability to control their orbit, their attitude, or both. According to ESA’s MASTER (Meteoroid and Space Debris Terrestrial Environment Reference) model at the reference epoch of 1 August 2024, approximately 54,000 objects larger than 10 cm are in orbit, including about 9,300 active payloads. In addition, an estimated 1.2 million objects between 1–10 cm and 130 million objects between 1 mm–1 cm populate the near-Earth space environment~\cite{ESA2025SpaceEnvironment}. The estimated number of debris objects larger than 1 cm, sufficient to cause catastrophic damage, exceeds one million~\cite{ESA2024Website}.

To address this growing risk, multiple agencies and research groups are actively exploring innovative strategies for space debris removal. Different concepts have been applied or discussed in the literature for ADR, including robotic arms, drag sails, harpoons, and nets~\cite{aglietti2020removedebris}. There are previous missions used to test ADR technologies, such as RemoveDEBRIS~\cite{aglietti2020removedebris}, which in 2018 successfully demonstrated ADR technologies using harpoons and nets, and the Kounotori Integrated Tether Experiment (KITE) by JAXA, which failed to deploy the tether due to a mechanical malfunction~\cite{ohkawa2020review}. Also, the Japanese Commercial Removal of Debris Demonstration (CRD2) which has two phases: Phase 1 involves non-cooperative rendezvous, proximity operations, and inspection (completed) ; Phase 2 is an ADR technology demonstration for a second-stage launch vehicle~\cite{nakamura2023study}. On the other hand, upcoming missions for ADR include ClearSpace-1 by ESA (planned for 2029), designed to remove in-space debris belonging to ESA and to demonstrate the technologies needed for future debris removal missions~\cite{biesbroek2021clearspace, ESA_ClearSpace1}.

 The existing approaches mainly rely on single-satellite servicers, which offer certain advantages but also come with several limitations and challenges. Therefore, this paper discusses the concept of swarm-based approach for active debris removal identifying key advantages and challenges of the concept. Table~\ref{tab:adr_comparison} summarizes the technologies used for ADR highlighting the expected advantages and disadvantages of each approach. 

\begin{table*}[!htbp]
\centering
\caption{Comparison of Active Debris Removal (ADR) and On-Orbit Servicing Approaches.}
\label{tab:adr_comparison}
\setlength{\extrarowheight}{2pt}
\footnotesize
\begin{tabularx}{\textwidth}{p{3cm} p{3cm} X X}
\Xhline{1pt}
\textbf{Approach} & \textbf{Example Missions} & \textbf{Advantages} & \textbf{Disadvantages} \\ \hline

Robotic Arms & ClearSpace-1~\cite{biesbroek2021clearspace,ESA_ClearSpace1} &
• High precision capture \newline
• Can handle large, irregular debris \newline
• Reusable for multiple captures if refueled &
• Mechanically complex \newline
• Requires accurate rendezvous \newline
• Single point of failure for the ADR technology \\ \hline

Harpoons & RemoveDEBRIS~\cite{aglietti2020removedebris} &
• Can engage debris from a safe distance \newline
• Simple capture mechanism &
• Limited to debris with penetrable surfaces \newline
• Single point of failure for the ADR technology \\ \hline

Nets & RemoveDEBRIS~\cite{aglietti2020removedebris} &
• Can capture tumbling debris \newline
• Simple and robust &
• Risk of entanglement with chaser \newline
• Difficult to control post-capture motion \newline
• Single point of failure for the ADR technology \\ \hline

Electrodynamic Tethers & Kounotori Integrated Tether Experiment (KITE)~\cite{ohkawa2020review} &
• No propellant required for de-orbit \newline
• Simple mechanism \newline
• Lightweight system &
• Limited to conductive environments (LEO) \newline
• Single point of failure for the ADR technology \\ \hline

Single-Satellite Propulsion Tug & Mission Extension Vehicle (MEV-1/2)~\cite{MEV12} &
• Proven rendezvous and docking with an active GEO satellite \newline
• Provides full propulsion and attitude control &
• Specific capturing/docking spot \newline
• Single point of failure for the ADR technology \\ \hline

Swarm of Small Satellites – NanoTugs (Proposed)~\cite{asri2024introductory, han2018integrated, udrea2015cooperative, matsuka2021collaborative, zhu2025bioinspired} & -- &
• Distributed redundancy — failure of one unit does not end mission \newline
• Multiple view for target inspection and characterization \newline
• Can adapt to debris shape via multiple attachment points \newline
• Scalable for different debris sizes \newline
• No single point of failure for the ADR technology; only the mother/carrier spacecraft is critical &
• Complex coordination and swarm control \newline
• Higher risk of in-orbit collision between swarm units \newline
• Requires robust communication and guidance algorithms \newline 
• Single point of failure for the mother spacecraft \\ \hline
\Xhline{1pt}
\end{tabularx}
\end{table*}

As shown in Table~\ref{tab:adr_comparison}, existing ADR approaches present unique strengths and limitations. Single-satellite missions, such as those employing robotic arms, nets and propulsion tugs, have demonstrated high precision and proven capture capabilities; however, they suffer from a single point of failure, where the loss of the key technology employed for the debris capturing results in the complete loss of mission objectives. On the other hand, the proposed swarm-based concept can offer distributed redundancy, enabling continued operations even in the event of individual unit failures especially when considering decentralized technologies between the agents. Moreover, the ability to attach multiple small satellites to various points on a debris object can provide greater flexibility in controlling attitude and distributing de-orbiting forces, which is particularly advantageous for large, irregular, or tumbling targets. The swarm concept also allows for scalability—deploying a feasible number of units based on debris characteristics—while leveraging standardized, potentially lower-cost platforms. However, this approach also comes with challenges. It requires complex coordination and swarm control, robust communication and guidance algorithms, and carries a risk of in-orbit collisions between the swarm units. Additionally, the mother spacecraft represents a single point of failure; however, a well-developed and demonstrated spacecraft, such as those demonstrated by D-Orbit’s ION service~\cite{dorbit_ion_service} or UniSat-7~\cite{unisat7}, can be used to mitigate this risk.

Asri and Zhu~\cite{asri2024introductory} provided a review of swarm technology for spacecraft on‐orbit servicing, including applications such as on‐orbit assembly and active debris removal. The authors discuss the advantages that a swarm-based approach can offer over single-platform solutions, while also highlighting the associated complexities and summarizing existing studies relevant to the concept. Although the use of spacecraft swarms for on‐orbit servicing is still rarely explored as a complete approach, considerable research has been conducted on its individual aspects, including swarm-based target inspection and characterization, the development of capturing mechanisms, and autonomous swarm guidance, navigation, and control (GNC)~\cite{asri2024introductory}.  

Target inspection and characterization are typically performed using sensors such as optical cameras and LiDAR to generate a 3D point cloud model of the target. Recently, artificial intelligence (AI) and machine learning (ML) techniques have also been introduced to estimate physical properties of the target, such as its mass and inertia tensor. Several studies have explored the use of multiple spacecraft for target characterization and pose estimation, highlighting their advantages over single-satellite approaches. Matsuka et al.~\cite{matsuka2021collaborative} demonstrated the benefits of a cooperative vision-based pose estimation algorithm that leverages multiple spacecraft to overcome the visibility limitations often faced by a single chaser. Their study showed that multi-agent, multi-view observations provide more accurate and efficient pose estimation of the target, whereas a single spacecraft would require significantly more time to obtain comparable data from different viewpoints.

In proximity operations and the capturing phase of targets, coordinated control is crucial to handle uncertainties such as unknown tumbling rates and limited observability of uncooperative objects. Traditional single-satellite solutions often lack the robustness needed in such scenarios, whereas swarm-based approaches provide enhanced redundancy, adaptability, and safety, where safety refers to minimizing the collision risks by distributing the interaction between the chasers and the target. For instance, Asri and Zhu~\cite{zhu2025bioinspired} presented a bio-inspired consensus-based control strategy in which multiple spacecraft cooperatively approach and capture an uncooperative target. Their approach leverages distributed consensus dynamics inspired by natural swarms, enabling autonomous decision-making, collision avoidance, and formation reconfiguration during proximity operations. Simulation results demonstrated that this strategy enhances robustness against target unpredictability and allows the swarm to efficiently coordinate sensing and actuation for a successful capture.

In the context of debris approach control, Kozin et al.~\cite{kozin2022laboratory} presented a laboratory test bed using unmanned aerial vehicle (UAV) mock-ups on a planar air-bearing platform to assess microsatellite control algorithms for active debris removal. The facility provides nearly frictionless two-dimensional motion, offering a close approximation of orbital dynamics and enabling experimental validation of different control strategies. The performance of the proposed control algorithms was evaluated using a debris mock-up with varying angular velocities. The results showed that both the virtual potential-based and state-dependent Riccati equation (SDRE)-based controllers successfully achieved docking at angular velocities below 12~deg/s, while neither algorithm succeeded at 18~deg/s. For successful docking cases, the virtual potential-based control required lower $\Delta V$ but took longer, as it waited for an acceptable relative position, whereas the SDRE-based control actively pursued docking conditions, resulting in shorter docking times but higher $\Delta V$. 

For ADR and, more broadly, on‐orbit servicing, the capturing or docking phase is one of the most critical stages of the mission. Since most target objects are not designed for docking, the capturing mechanism must be capable of adapting to various surface properties. Additionally, the mechanism must withstand the impact forces generated during capture and provide sufficient holding strength to maintain attachment until the mission is completed. Several researchers have investigated potential technologies for robotic attachment with ground testing in zero-gravity environments~\cite{spenko2023making}. One promising material is the gecko-like adhesive, which has gained significant attention for space applications due to its ability to cling to both flat and curved surfaces and its reliable performance in vacuum~\cite{spenko2023making}. Delisle et al.~\cite{hubert2023hybrid} proposed a hybrid ADR system that integrates controlled active linear actuators with passive gecko adhesives, enabling gentle yet secure contact between the chaser and the target. Ben-Larbi et al.~\cite{ben2022orbital} highlighted adhesive-based capture techniques, particularly micropatterned dry adhesive surfaces that rely on van der Waals forces, as a promising solution due to their simplicity and suitability for the space environment. The authors examined various material types and target scenarios, including fixed and free-floating objects, as well as surfaces with different properties, curvatures, and conditions. They pointed out that although extensive laboratory studies have advanced the understanding of adhesion mechanisms on fixed substrates, these findings cannot be directly applied to ADR missions involving free-floating targets with six degrees of freedom. The main challenges identified include sensitivity to misalignment, collision dynamics, compressive load limitations, and the effects of dust contamination and surface degradation on target objects. To overcome these limitations, the authors emphasize the need for quantitative investigations under realistic operational conditions to assess the performance boundaries and practical applicability of micropatterned adhesives in ADR missions. The authors also emphasized the advantage of distributing the capture points to better account for the curvature of the target surface, rather than relying on a single attachment point. This further strengthens the case for the swarm-based approach proposed in this paper, as multiple gripping points enable a better interaction with irregularly shaped debris, something that might be challenging for a single servicer spacecraft.

Post-capturing, only a few studies have investigated the visibility and task allocation of different objectives for swarm agents. Hana et al.\cite{han2018integrated} proposed an approach where nanosatellites are attached to a failed satellite to take over its attitude control functionality. Using real-time estimation of the combined structure’s inertia matrix, an attitude takeover controller is designed with appropriate task allocation for each nanosatellite. Similarly, Udrea and Nayak\cite{udrea2015cooperative} presented an overview and preliminary design of a cooperative multi-satellite mission for controlled ADR in Low Earth Orbit. Their concept involves a mother spacecraft carrying six 12U CubeSats, which are deployed in proximity to an upper-stage rocket body to dock using an active electrostatic adhesive mechanism. The nanosatellites, equipped with thrusters and reaction wheels, are employed to de-tumble the target, preparing it for the de-orbiting phase, after which the mother spacecraft docks to the stabilized upper-stage for de-orbiting. Net-based debris capture mechanisms have also been investigated as an alternative approach for cooperative debris removal. In these concepts, a flexible net equipped with multiple actuated nodes is deployed from a servicing spacecraft to capture the debris \cite{ru2022capture, zhao2020capture}. Although such systems involve multiple actuators, they are generally modeled as a mechanically coupled structure rather than a swarm of independent agents, since the nodes are physically connected through tethers and typically deployed from a single spacecraft. Nevertheless, these concepts highlight the potential advantages of distributed capture mechanisms and coordinated actuation for handling non-cooperative tumbling debris.

The discussed approaches for ADR using a swarm of nanosatellites have not examined the use of the swarm throughout all mission phases. For example, ~\cite{udrea2015cooperative} considered the swarm for attitude control only, while the de-orbiting was handled by the mother spacecraft. To address this gap, the present work explores an innovative concept where a swarm of nanosatellites autonomously de-orbits large debris objects, such as rocket bodies. The primary goal of this paper is to present an overview of the concept along with a preliminary nanosatellite design, analysis of swarm size to de-orbit debris objects of various masses at different orbit altitudes, focusing mainly on the debris stabilization and de-orbiting phase in terms of control and task allocation. It should be emphasized that ADR technologies and concepts are still in the development and testing stages, both in laboratories and through experimental missions. Therefore, single-satellite ADR approaches are regarded as initial attempts, while swarm-based concepts will only be considered once the technologies become more reliable and sufficiently mature. 

The paper is structured as follows. Section~\ref{sec:conceptOverview} presents the swarm-based ADR mission concept, including the mission phases, the definition of mission and system parameters, and the analytical approach for system sizing along with a preliminary NanoTug design. Section~\ref{sec:numerical} details the numerical study, covering NanoTug distribution on the debris, system dynamics, and control laws for stabilization and de-orbiting, as well as example mission scenarios that connect to the analytical approach. Finally, Section~\ref{sec:Conclusion} provides concluding remarks and discusses possible directions for future work.

\section{Mission Concept and Preliminary Design Analysis} 
\label{sec:conceptOverview}

\subsection{General Mission Overview}
This paper investigates the potential of a swarm-based approach for active space debris removal, building upon the work presented in~\cite{Alnaqbi2025_SwarmDrivenSolution} with additional analyses, considerations, and a thorough discussion of the current status and directions for future studies. In this concept, a mother spacecraft deploys a set of nanosatellites (referred to as NanoTugs) to conduct debris removal missions. The mother spacecraft provides the primary maneuvering capability for large orbital transfers to the targeted debris. Following the planned ejection and commissioning activities, it releases a swarm of the NanoTugs that autonomously execute the de-orbiting mission. Initially, the NanoTugs, each equipped with five thrusters mounted along different axes, establish a distributed bounded motion around the debris, enabling detailed characterization through vision-based navigation. A decentralized communication approach is considered between the agents to account for the obstruction caused by the debris in this phase. Working collaboratively, the swarm builds a comprehensive understanding of the debris' physical and dynamic properties, facilitating the selection of optimal capture points to achieve full-state control.

Each NanoTug is assumed to be equipped with a gecko-inspired capturing mechanism designed to absorb the initial impact and maintain a secure attachment to the debris throughout the mission. The NanoTugs are also equipped with active attitude control system based on reaction wheels, enabling them to align to the required attitude for gripping. To account for reaction wheels saturation, magnetorquers can be used. Once the debris is captured, the swarm proceeds with the de-tumbling phase followed by the de-orbiting maneuver using coordinated thruster operations in both, which will transfer the debris to a low orbit altitude based on mission requirements. The reaction wheels are not assumed to be used in the de-tumbling and de-orbiting phase. As the NanoTugs remain attached to different points on the debris, the decentralized communication and control architecture is employed to enable autonomous decision-making and task allocation. This system determines which NanoTugs should activate their thrusters for de-tumbling and for de-orbiting and which should manage attitude control during the de-orbiting. Depending on the mission profile, the NanoTugs can be reused for subsequent debris-removal missions by re-attaching to the mother spacecraft. The mother spacecraft can follow the de-orbiting trajectory that enables NanoTug re-attachment and may also be equipped with refuelling capabilities to prepare the NanoTugs for subsequent de-orbiting missions.\\

\noindent This concept is divided into six phases, where the process is iterative after each debris removal mission (if applicable):

\begin{enumerate}
    \item \textbf{Orbital Transfer and NanoTugs Release}: the mother spacecraft performs an orbital transfer and deploys a number of NanoTugs in the proximity of the target debris based on a preliminary inspection of the debris properties.
    
    \item \textbf{NanoTugs Swarm Establishment}: the NanoTugs establish a bounded and coordinated motion around the debris to begin observation.

    \item \textbf{Inspection and Characterization}: the NanoTugs use the payload sensors to identify precisely the physical properties of the debris.
    
    \item \textbf{Capturing}: the NanoTugs initiate and execute the debris capturing activities.
    
    \item \textbf{Stabilization and De-Orbiting}: the captured debris is stabilized (de-tumbled) and transferred to a low altitude orbit using NanoTugs.
    
    \item \textbf{Re-Docking with the Mother Spacecraft}: The NanoTugs detach from the debris and re-dock with the mother spacecraft, which follows the de-orbiting trajectory to facilitate re-attachment. This phase is executed only for NanoTugs with sufficient remaining propellant to complete the return maneuver. The NanoTugs that perform the re-docking can then be refuelled by the mother spacecraft for subsequent debris-removal missions.
\end{enumerate}

\begin{figure*}[!h] 
\renewcommand{\arraystretch}{1.5}
    \centering
        \includegraphics[width=1\textwidth]{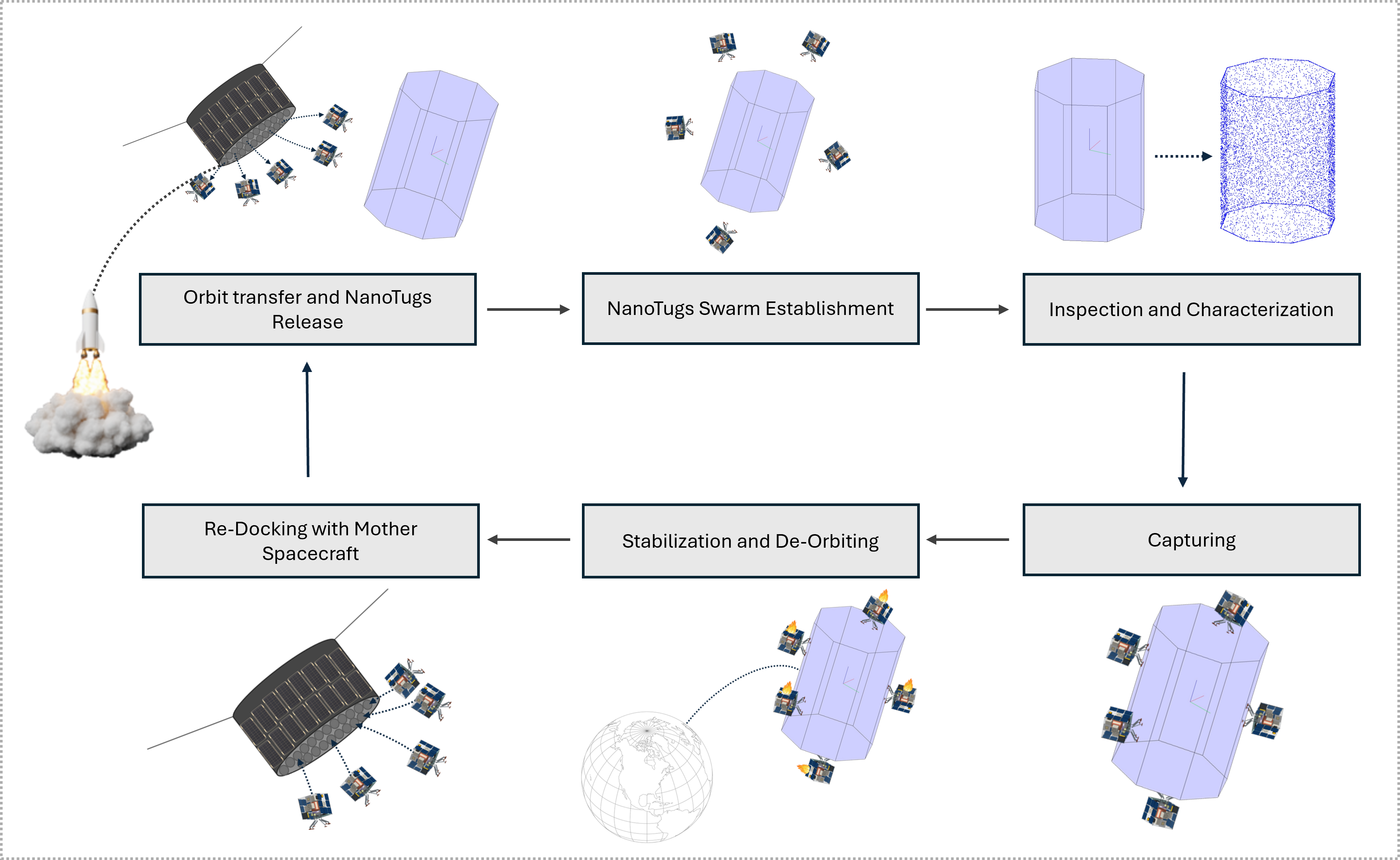}
    \caption{Mission Concept Overview.}
    \label{fig:conceptOverview}
    
\end{figure*}

Figure~\ref{fig:conceptOverview} illustrates the proposed mission concept. This paper focuses on the fifth phase, \textbf{Stabilization and De-Orbiting} of the debris, while the other phases will be addressed in future studies. It is assumed that phases 1–4 have already been completed, and the debris has been successfully captured by the NanoTugs.

The work focuses on the preliminary design of such a system, requiring the derivation of feasible NanoTug characteristics and the determination of the swarm size for a given debris de-orbiting mission. The ADR system design is supported by numerical modeling of the coupled 6-DoF (translational and rotational) dynamics and control of a debris object with attached NanoTugs, treated as a single rigid body.

\subsection{ADR System Preliminary Design}
This section describes the relationship between mission parameters and the design parameters of an ADR system based on a swarm of NanoTugs. The objective is not to define a fixed system configuration, but rather to provide a general framework that supports understanding of the design approach, enabling its adaptation to a wide range of mission requirements and constraints. It is important to note that the analysis presented here is a preliminary assessment of mission performance for the selected system. More accurate and detailed evaluations should be carried out through numerical simulations, examples of which are presented later in this paper.\\
The parameters considered in this study are grouped into two main categories: \textit{mission parameters} and \textit{ADR system parameters}.\\ \\
\noindent \textit{Mission parameters} define the properties of the target debris and the operational constraints:
\begin{itemize}
\item Debris physical properties: mass $M_d$, inertia tensor $J_d$, and geometry (size and shape), from which the effective surface area $A_d$ is derived
\item Initial orbit altitude: $h_0$
\item Disposal orbit altitude: $h_f$
\item Expected mission duration: $t_\text{expected}$
\end{itemize}
\medskip
\noindent \textit{ADR system parameters} describe the characteristics and configuration of the NanoTug swarm:
\begin{itemize}
\item NanoTug properties: dry mass $M_{i_{\text{dry}}}$, initial propellant mass $M_{i_{\mathrm{prop},0}}$, specific impulse $I_{sp}$, thrust $T_i$, and characteristic dimensions (height $H$, width $W$, and length $L$)
\item Total swarm size: $N$ total number of NanoTugs attached to the debris
\item NanoTugs utilization ratio: $\lambda = \frac{n}{N}$, where $n$ is the average number of NanoTugs that are simultaneously used for de-orbiting. $\lambda$ highly depends on the distribution of the NanoTugs on the debris surface and the considered NanoTugs' thrusters task allocation strategy.  
\end{itemize}

\noindent\textit{Disposal Orbit Altitude Considerations} \\
\indent The disposal orbit altitude requirements come from mission parameters. The selection of the disposal orbit altitude depends on the desired orbital lifetime after the active de-orbiting maneuver, $t_{\text{expected}}$. According to ESA, this post-mission lifetime should not exceed five years for objects in low Earth orbit~\cite{ESA_ZeroDebris_2023}. After reaching the disposal orbit, the debris undergoes passive de-orbiting due to atmospheric drag at low altitudes. This phase corresponds to the natural decay of the orbit, ultimately leading to re-entry, and defines the lifetime of the debris following the active removal phase. The lifetime of the debris depends on the debris ballistic coefficient which is a function of debris of mass $M_d$, drag coefficients $C_d$ and surface area $A_d$ defined as follows~\cite{LarsonWertz1999}:

\begin{equation}
BC = \frac{M_d}{C_d A_d}
\end{equation}

\begin{figure}[h!]
    \centering
        \includegraphics[width=0.9\textwidth]{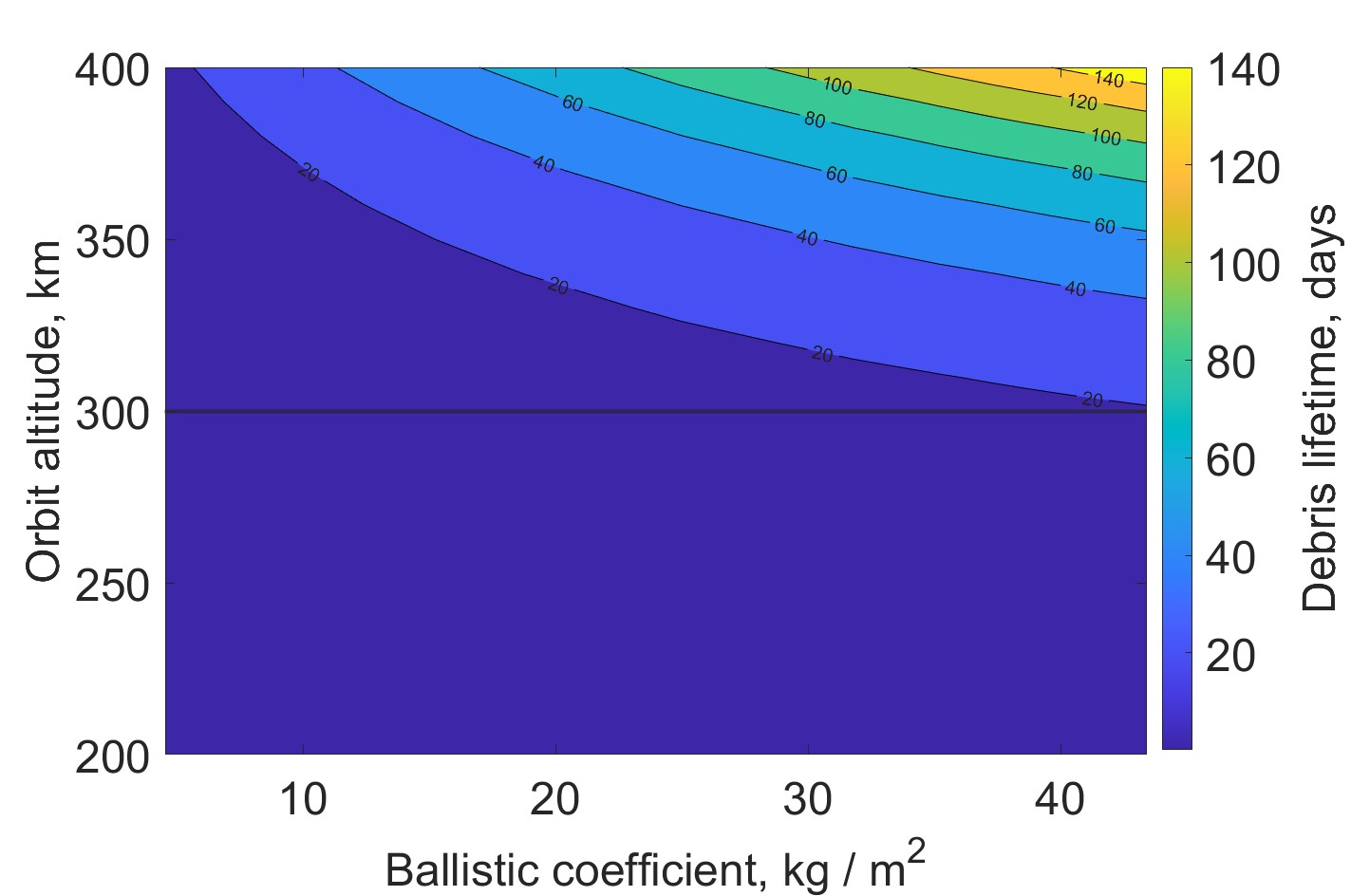}
    \caption{Debris Lifetime.}
    \label{fig:DebrisLifeTime}
\end{figure}

Figure~\ref{fig:DebrisLifeTime} shows the lifetime of debris with different ballistic coefficient and different disposal orbit. The lifetime is calculated via numerical orbit propagation in the two-body problem with the conventional atmospheric drag model using piece-wise exponential model~\cite{vallado2001fundamentals}. Selecting a higher disposal orbit reduces the propellant requirements of the ADR system, potentially lowering the number of required NanoTugs and enabling faster active de-orbiting operations. Additionally, it makes it easier to raise the NanoTugs’ orbit afterward, if needed for another de-orbiting mission. However, placing the debris at a higher altitude (around 400 km) introduces additional risks due to the higher density of operational satellites in that region~\cite{ESA2025_SpaceEnvironment}. A lower disposal altitude, on the other hand, benefits from a reduced number of active satellites at that altitude, but requires more propellant, a larger number of NanoTugs, and longer maneuver durations; it also makes post-mission orbit raising of NanoTugs more challenging. Nevertheless, the passive de-orbiting becomes faster. For example at 300 km altitude and for the range of debris ballistic coefficients considered, Fig.~\ref{fig:DebrisLifeTime} indicates that the debris would de-orbit to Karman line, 100 km altitude, in less than 20 days.\\

\noindent\textit{Debris De-Orbiting Analysis} \\
\indent Debris de-orbiting analysis enables the assessment of the minimum required number of NanoTugs and the time needed to de-orbit a debris of a given mass from its initial altitude to the disposal orbit. The analysis does not account for the propellant consumed during debris stabilization and assumes idealistic control authority of the ADR system. Therefore, this analysis provides a lower bound on both the number of NanoTugs required and the time needed to reduce the debris orbit altitude to the disposal orbit.

The time required to lower a circular orbit from an initial radius $r_0 = h_0 + R_\oplus$ to a final radius 
$r_f = h_f + R_\oplus$ ($R_\oplus$ is the mean radius of the Earth) under continuous low-thrust can be derived using Gauss’ variational equations~\cite{battin1999introduction}. Assuming a circular orbit (a $\approx$ r) and considering a constant thrust in the direction opposite to the velocity vector, the change of the semi-major axis $a$ is governed by

\begin{equation}\label{eq: sma rate, GVE}
    \frac{da}{dt} = -\frac{2T_S}{M_S(t)}\frac{a^{3/2}}{\sqrt{\mu}} ,
\end{equation}

\noindent where $T_S = nT_i$ is the total thrust of the ADR system, assuming n NanoTugs use thrusters in the required direction, and $M_S(t)$ is the time-varying debris-NanoTugs system mass due to propellant consumption and $\mu$ is the Earth gravitational parameter.
The initial mass of the debris-NanoTugs system $M_{S,0}$ and its time derivative are follows:

\begin{subequations}

\begin{equation}\label{eq:system initial mass}
M_{S,0} = M_d + N (M_{i_\mathrm{dry}} + M_{i_\mathrm{prop,0}}),    
\end{equation}

\begin{equation}\label{eq:system mass rate}
\dot M_S = -\frac{T_S}{I_{sp} g_0}.\\
\end{equation}

\end{subequations}
\noindent where $g_0$ is the standard gravitational acceleration.\\
\noindent Taking into account the equations~\ref{eq:system initial mass}, \ref{eq:system mass rate} the system mass changes linearly and system mass as a function of time $M_S(t)$ is as follows:

\begin{equation}\label{eq:mass as time}
M_S(t) = M_{S,0} - \frac{T_S}{I_{sp} g_0} t. \\    
\end{equation}
Substituting~\ref{eq:mass as time} into~\ref{eq: sma rate, GVE}, separating variables and defining limits of integration yields

\begin{equation}
    \int_{r_0}^{r_f} a^{-3/2} \, da = -\frac{2 T_s}{\sqrt{\mu}M_{S,0}} \int_0^{t_\text{de-orb}} \frac{dt}{1 - \frac{T_s}{I_{sp}g_0M_{S,0}} t}.
\end{equation}
Performing the integration and substituting integration limits give

\begin{equation}
    -\frac{2}{\sqrt{a}} \Big|_{r_0}^{r_f}
    =  -\left(-\frac{2 I_{sp}g_0}{\sqrt{\mu}} \ln(1 - \frac{T_s}{I_{sp}g_0M_{S,0}} t) \Big|_0^{t_\text{de-orb}}\right),
\end{equation}

\begin{equation}\label{eq:integral solution}
    2 \left( \frac{1}{\sqrt{r_0}} - \frac{1}{\sqrt{r_f}} \right)
    = \frac{2 I_{sp}g_0}{\sqrt{\mu}} \ln(1 - \frac{T_s}{I_{sp}g_0M_{S,0}} t_\text{de-orb}).
\end{equation}
Let us solve the Eq.~\ref{eq:integral solution} for $t_\text{de-orb}$ to determine the time required to lower a circular orbit from radius $r_0$ to $r_f$, taking into account the limited propellant available for the maneuver and the continuous use of $n$ thrusters within the swarm. The latter is important because not all NanoTugs can provide control input in the desired direction at all times. In this analysis, atmospheric drag is neglected:

\begin{subequations}
\label{eq: descent time both}
\begin{align}
t_\text{de-orb} &= \frac{I_{sp} g_0 \left[ M_d + N \left( M_{i_\mathrm{dry}} + M_{i_\mathrm{prop,0}} \right) \right]}{n T_i}
\left[ 1 - \exp\left(-\frac{\sqrt{\mu}}{I_{sp} g_0}
\left( \frac{1}{\sqrt{r_f}} - \frac{1}{\sqrt{r_0}} \right) \right) \right],  \label{eq: descent time, system sizing}\\[3pt]
t_\text{de-orb} &\leq t_\text{control}^{\max},
\label{eq: limit on de-orbiting time}
\end{align}
\end{subequations}

\noindent where $t_\text{control}^{\max}$ is the maximum time the swarm can perform de-orbiting given the available propellant if using $n$ thrusters simultaneously 
\begin{equation}\label{eq: maximum descent time for a given propellan}
t_\text{control}^{max} = \frac{N}{n}\frac{ I_{sp} g_0 M_{i_\mathrm{prop,0}}}{T_i} = \frac{1}{\lambda}\frac{ I_{sp} g_0 M_{i_\mathrm{prop,0}}}{T_i}.
\end{equation}
Equations~\ref{eq: descent time, system sizing}, \ref{eq: limit on de-orbiting time} allow us evaluate de-orbiting time for a given mission parameters depending on the swarm size and NanoTugs propulsion characteristics and thus can be used for the ADR system sizing.\\

The evaluation of mission performance requires the initial identification of the NanoTugs' propulsion system properties $I_{sp}$ and $T_i$. The selection of these parameters is often iterative to provide the optimal solution for a given mission. The selection of the thruster parameters complies with the performance limits of existing thruster technologies for small satellites. Studies such as~\cite{alnaqbi2024propulsion, parker2016state} provide a comprehensive overview of available propulsion systems for small satellites and highlight the differences in their characteristics. In this context, a trade-off must be considered: selecting a high $I_{sp}$ reduces propellant consumption, allowing for smaller tanks, lighter NanoTugs, and fewer units required per mission; whereas selecting a high thrust ensures that the de-orbiting mission can be completed within a reasonable time frame. Referring to Fig.~38 in~\cite{alnaqbi2024propulsion}, bipropellant, monopropellant, and resistojet systems show a reasonable range of specific impulse $I_{sp} \in [90, 400]$~s and a practical range of thrust $T_i \in [0.1, 10]$~N.

\begin{figure*}[!h] 
\renewcommand{\arraystretch}{1.5}
    \centering
        \includegraphics[width=1\textwidth]{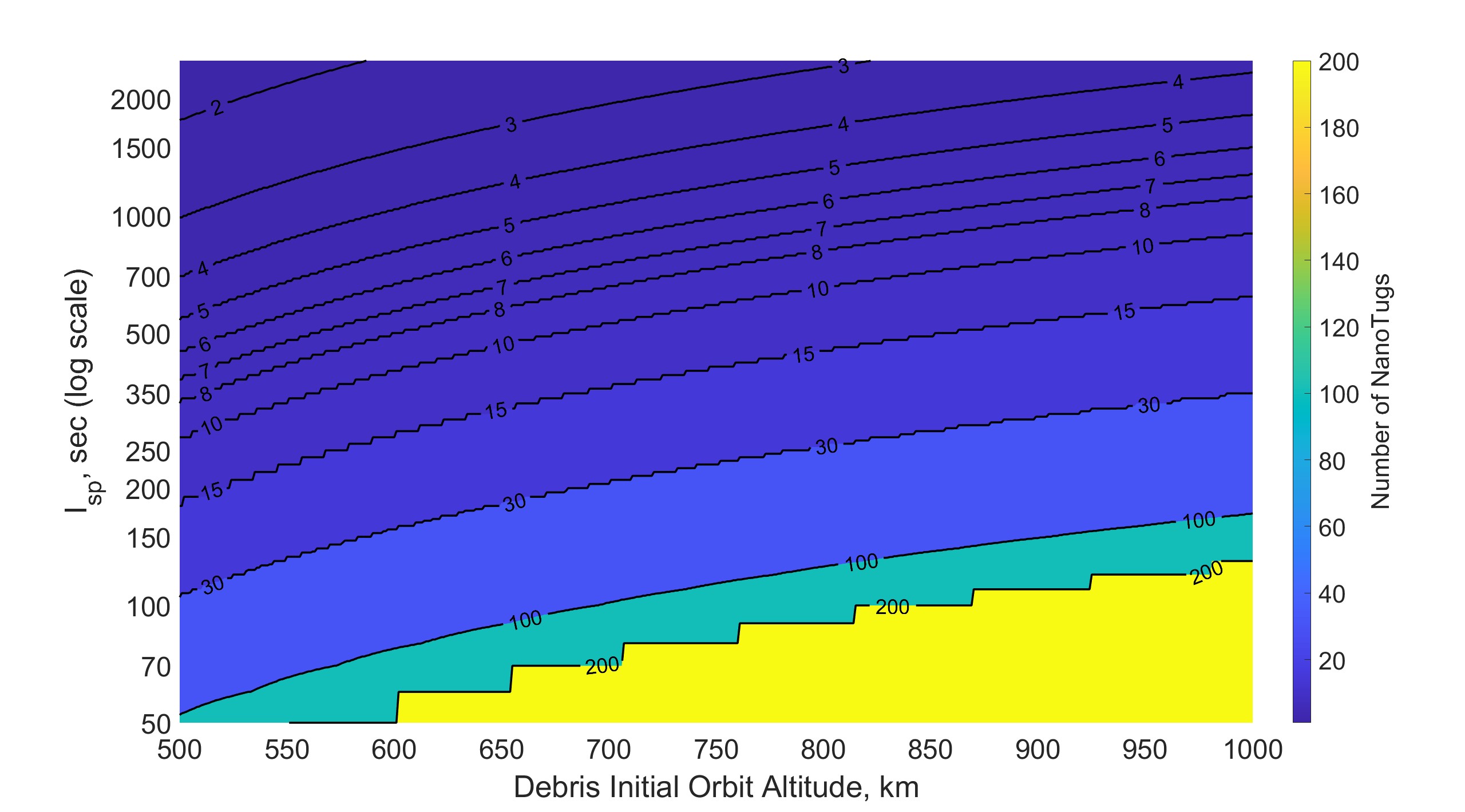}
    \caption{Number of NanoTugs Required for De-orbiting using 5 kg Propellant Mass with a Range of Initial Altitude and Isp, Debris Mass = 1000 kg.}
    \label{fig:ispVsAltitudeVsN1000kgDebris}
    
\end{figure*}

To provide an overview of the relationship between mission parameters and system parameters, Eqs.~\ref{eq: descent time both} and~\ref{eq: maximum descent time for a given propellan} can be used by fixing some parameters and varying others. For example, assuming $\lambda = 1$, which implies that all thrusters in the swarm system are used simultaneously during the mission ($n = N$), and by equating Eq.~\ref{eq: descent time, system sizing} with Eq.~\ref{eq: maximum descent time for a given propellan}, the thrust term $T_i$ cancels out. This yields an expression that depends on the system masses, propulsion properties, and orbital parameters, while being independent of the thrust magnitude and de-orbiting time. This approach assumes that the debris de-orbiting will occur within the maximum time the swarm system can operate before exhausting all propellant, $t_\text{control}^{\max}$. This method allows the determination of the minimum required number of NanoTugs as a function of $I_{sp}$ and $r_0$ for a given set of constant parameters.

\begin{equation}
    n = \arg \min_{n \in \mathbb{Z}^+} \left| t_\text{de-orb}(n) - t_\text{control}^{max} \right|.
    \label{eq:numberOfNanoTugs}
\end{equation}
Figures~\ref{fig:ispVsAltitudeVsN1000kgDebris} illustrates the resulting minimum number of NanoTugs required for a debris mass of 1000 kg as a function of $I_{sp}\in [50, 2500]$~s and $h_0 \in [500, 1000]$~km. The disposal orbit altitude is set to $h_f = 300$~km. The NanoTugs dry mass is set to $M_{i_{\text{dry}}} = 10$ kg and each NanoTugs carry propellant of mass $M_{i_\mathrm{prop,0}} =5$ kg. The maximum number of NanoTugs considered in solving Eq.~\ref{eq:numberOfNanoTugs} is 200. As shown in the figure, selecting a specific impulse $I_{sp} < 100$~s results in significantly higher requirements in terms of the number of NanoTugs or propellant mass.\\

\begin{figure*}[!h] 
\renewcommand{\arraystretch}{1.5}
    \centering
        \includegraphics[width=1\textwidth]{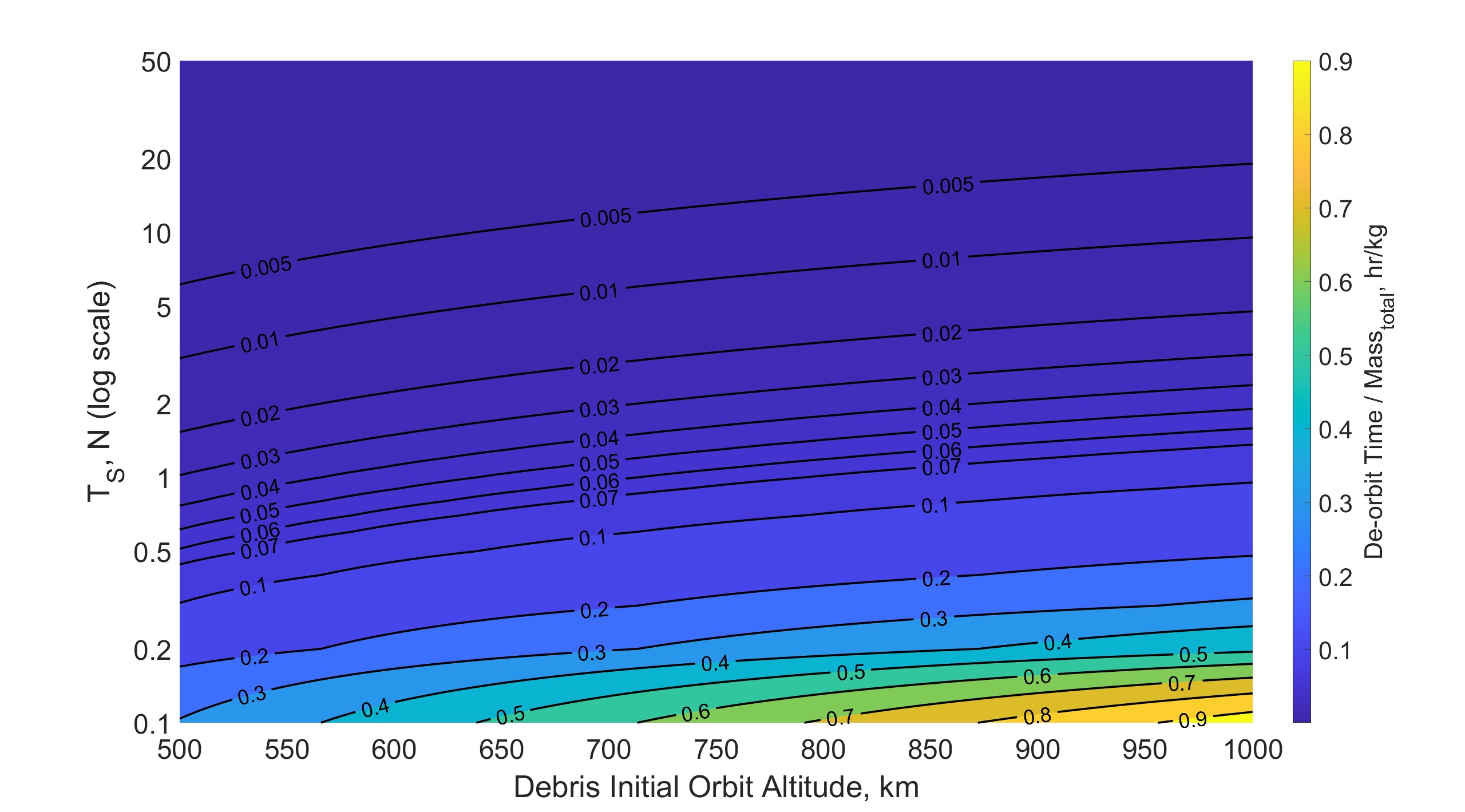}
    \caption{De-orbit Time per System Total Mass for a Range of Initial Debris Altitude and Total applied Thrust, $I_{sp} = 200$~sec.}
    \label{fig:thrustVsAltitudeVstimePerMass}
    
\end{figure*}

On the other hand, Eq.~\ref{eq: descent time, system sizing} can be used to find the time required for de-orbiting per initial system mass $\frac{t_\text{de-orb}}{M_{S,0}}$ for a range of total system thrust $T_S = nT_i$, a range of $h_0$ and for constant $I_{sp}$ and $h_f$. Figure~\ref{fig:thrustVsAltitudeVstimePerMass} illustrates the de-orbit time per unit system mass for a range of initial debris altitudes and total applied thrust values, $I_{sp} = 200 \, \text{sec}$ and $r_f = 300$~km. \\

\noindent\textit{Debris De-tumbling Analysis}\\
\indent Based on~\cite{vsilha2020space} the observation of various types of debris shows that the apparent angular velocity of the debris such as SL-12 rocket bodies~\cite{SL12RB15338} can reach up to $80 \, \text{deg/s}$. To account for the debris tumbling in the calculation of the required number of NanoTugs and the de-orbit time, the propellant mass necessary for de-tumbling and time required to de-tumble is considered. The estimate is intended only as a first-order, order-of-magnitude estimate for preliminary sizing. It is based on simplified assumptions and is not meant to represent the full 3D post-capture dynamics. To assess whether the predicted trends remain reasonable under more realistic conditions, the de-tumbling phase is further evaluated numerically using randomly generated initial rotational states and the coupled rigid-body dynamics model.

For a first-order analytical estimate, we assume that the debris initially rotates with an angular velocity $\omega_0$ along one of its principal axes of inertia with the maximum moment of inertia $J_S$ and the applied control torque $\tau$ is constant and has a direction opposite to the angular velocity vector. This allows to omit consideration of the gyroscopic term in the Euler equations of rotational motion and let us consider scalar equation for detumbling time and propellant consumption analysis:

\begin{equation}\label{eq: Euler simplified}
    J_{\text{S}} \, \dot{\omega} = \tau,
\end{equation}
The control torque $\tau$ from the ADR system depends on the shape and size of the debris, and NanoTug distribution among the surface of the debris. For the purpose of first-order estimate on the propellant consumption for debris detumbling let us consider an average control torque authority of the system using the control torque $\tau$ in the form of

\begin{equation}\label{eq:control torque assumption}
    \tau = R_{\text{eq}} \ T_i,
\end{equation}

\noindent where $R_{\text{eq}}$ is the equivalent moment arm for the thrust. The equivalent  moment arm is chosen such that a debris with a cylinder shape of height $H_d$ and radius $R_d$ has the same volume as a spherical debris of radius $R_{\text{eq}}$
\begin{equation}
R_{\text{eq}} = \left(\frac{3}{4} R_d^2 \, H_d \right)^{1/3}
\end{equation}
The required propellant to de-tumble the debris can be calculated as follows

\begin{equation}
M_{\text{prop}} 
= \int_0^{t_\text{detumble}} \dot{M}_{\text{prop}} \, dt
= \int_0^{t_\text{detumble}} \frac{T_i}{I_{\text{sp}} \, g_0} \, dt
= \frac{J_{\text{S}} \, \omega_0}{R_{\text{eq}} \, I_{\text{sp}} \, g_0}.
\end{equation}
On the other hand, the time required to de-tumble the debris is 
\begin{equation}
t_\text{detumble} = \int_0^{t_\text{detumble}} dt = \int_{\omega_0}^{0} \frac{d\omega}{\dot{\omega}} = \int_{\omega_0}^{0} \frac{J_{\text{S}}}{R_{\text{eq}} \, T_i} \, d\omega = \frac{J_{\text{S}} \, \omega_0}{R_{\text{eq}} \, T_i}.
\end{equation}

Figure~\ref{fig:ispVsOmegaVsProp} illustrates the propellant mass required to de-tumble a debris object as a function of specific impulse and initial angular velocity along the axis with the biggest moment of inertia, while Fig.~\ref{fig:detumblingTime} shows the corresponding de-tumbling time with a single NanoTug thrust of $T_i$. The debris is assumed to be cylindrical, with mass $M_d = 1000$~kg, height $H_d = 10$~m, and radius $R_d = 3$~m. Torque is applied at a lever arm $R_{\text{eq}}$, and the debris moment of inertia is taken as the maximum about the principal axes, calculated as 

\begin{equation}
    J_{\text{S}} = \frac{1}{12} M_d \left( 3 R_d^2 + H_d^2 \right).
\end{equation}
As seen in Fig.~\ref{fig:ispVsOmegaVsProp}, the propellant required to de-tumble the debris is small compared to that needed for de-orbiting. For the range of specific impulse and initial angular velocities below 50~deg/s, less than 5~kg of propellant is necessary, indicating that a single NanoTug with 5~kg of propellant is sufficient for this task. 
Furthermore, Fig.~\ref{fig:detumblingTime} shows that the de-tumbling time using a single NanoTug with the given thrust is on the order of a few hours, especially for thrusts of 0.3~N and above, and for the considered range of initial angular velocities. In actual mission scenarios, multiple NanoTugs can operate simultaneously, so the de-tumbling time is expected to be shorter than the values predicted by this single-NanoTug analytical estimate.

\begin{figure*}[!h] 
\renewcommand{\arraystretch}{1.5}
    \centering
    \includegraphics[width=1\textwidth]{ispVsOmegaVsProp.jpg}
    \caption{Propellant mass required to detumble a cylindrical debris object as a function of specific impulse and initial angular velocity along the axis with the biggest moment of inertia.}
    \label{fig:ispVsOmegaVsProp}
\end{figure*}

\begin{figure*}[!h] 
\renewcommand{\arraystretch}{1.5}
    \centering
    \includegraphics[width=1\textwidth]{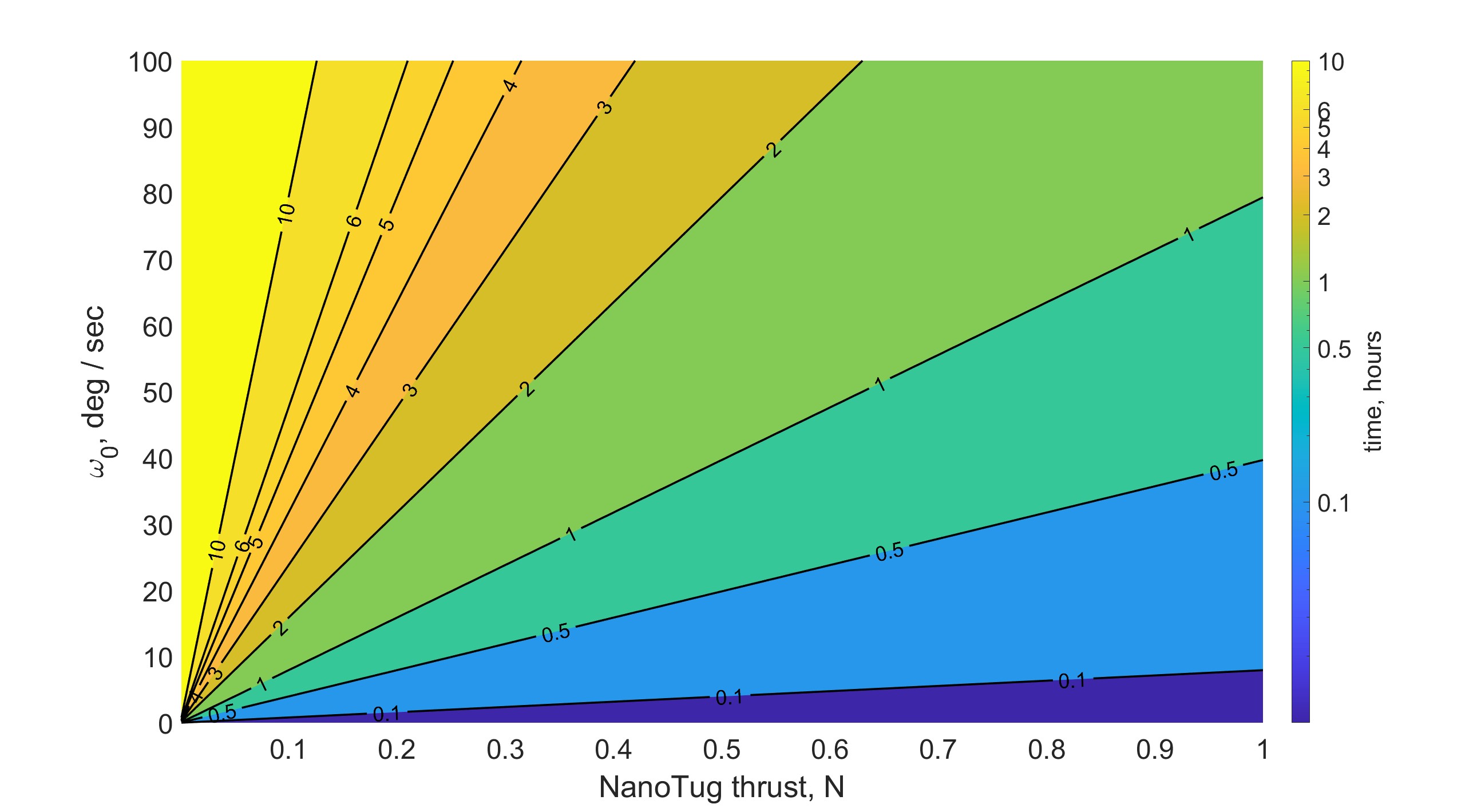}
    \caption{De-tumbling time for a 1000~kg cylinder as a function of initial angular velocity $\omega_0$ and applied NanoTug thrust.}
    \label{fig:detumblingTime}
\end{figure*}
The analysis emphasizes that the propellant requirement and time needed for de-tumbling are small compared to those for de-orbiting. Therefore, the estimates from de-orbiting can be used as a first-order approximation, with additional margin in propellant, number of NanoTugs, and mission time to account for the relatively small requirements of de-tumbling. It should be noted, however, that detailed numerical analysis is necessary for rigorous mission analysis.\\

\noindent\textit{System Sizing Approach} \\
\indent For a certain de-orbiting mission, the sizing of the swarm-based ADR system and the evaluation of mission performance follows the next steps. The proposed NanoTug concept is intended for large debris objects, such as upper stages of launch vehicles, which provide sufficient surface area for multiple NanoTugs to attach and perform controlled de-orbiting. Representative masses for these targets are $\sim$ 50--4000~kg, consistent with tracked rocket body debris ~\cite{Anselmo2016_RankingUpperStages}. On the other hand, the NanoTugs are assumed to follow nanosatellite requirements, so their mass and volume should be within the typical NanoSat/CubeSat standard limits, with a wet mass of $\leq 15$~kg (allowing for a small design margin) and a volume of $\leq 8$--12~U~\cite{nasa_cds_rev14_1}.

\noindent The system sizing approach is as follows, it assumes that all NanoTugs will consume all the propellant and the debris will de-orbit in  $t_\text{expected}$, where $n$ NanoTugs are operating simultaneously during the mission (example of using the approach will be shown later in the paper):

\begin{enumerate}
    \item Define mission parameters: define $M_d$, $h_0$, $h_f$ and $t_\text{expected}$\\
    
    \item Define candidate system parameters: specify $M_{i_\mathrm{dry}}$, $I_{sp}$, $T_i$, and $\lambda$. The chosen $I_{sp}$ and $T_i$ should reflect the performance limits of existing thruster technologies~\cite{alnaqbi2024propulsion}, while $\lambda$ depends on the NanoTugs’ distribution over the debris surface\\
    
    \item Compute required propellant: for the selected combination of $\lambda$, $I_{sp}$, and $T_i$, and by setting $t_\mathrm{control}^{\max} = t_\text{expected}$, calculate the required propellant per NanoTugs $M_{i_\mathrm{prop,0}}$ to operate for $t_\text{expected}$ using Eq.~\ref{eq: maximum descent time for a given propellan}\\
    
    \item Verify feasibility: check if $M_{i_\mathrm{prop,0}}$ is reasonable for the expected NanoTugs mass and volume budget. If not adjust one or more parameters defined in step 2 and repeat step 3\\
    
    \item Determine number of NanoTugs: once step 4 is satisfied, calculate $n$ required to satisfy $t_\mathrm{de-orb} = t_\text{expected}$ for the given debris parameters in step 1 and system parameters defined in step 2. The value is rounded up to the nearest integer using the ceiling function\\
    
    \item Calculate total number of NanoTugs: given the selected $\lambda$ and the calculated $n$, determine the total number $N$ of NanoTugs required to de-orbit the debris within $t_\mathrm{de-orb} = t_\text{expected}$. The value is rounded up to the nearest integer using the ceiling function
\end{enumerate}
The process should be iterative to find a feasible solution for the ADR system sizing problem. For example $t_\text{expected}$ can be smaller which will result in smaller $M_{i_\mathrm{prop,0}}$ in step 3 but will give higher number of NanoTugs in step 5 and 6. \\

\subsection{NanoTug Sizing}
Providing an exact design and sizing of the NanoTugs for the swarm is not the objective of this paper. The NanoTugs should be designed based on mission requirements and highly depend on the propulsion system selection and the propellant mass per NanoTug, which can occupy the largest budget of mass and volume. The NanoTugs should include all the major subsystems required, such as an on-board computer, communication system, electrical power system with batteries and solar panels, and the relative GNC subsystem to enable the rendezvous and mating phases with the target debris. An attitude determination and control system (ADCS), including reaction wheels, is required and will be used during docking with the debris and other mission phases. For the debris inspection phase, a camera is needed to characterize the debris using the NanoTugs. One of the important elements is the gripping mechanism, which should be designed to enable the NanoTugs to grasp the debris even when it is tumbling and to maintain a strong attachment throughout the mission. The exact design and selection of these subsystems are beyond the scope of this paper. 

\begin{figure}[!h]
    \centering
        \includegraphics[width=0.8\textwidth]{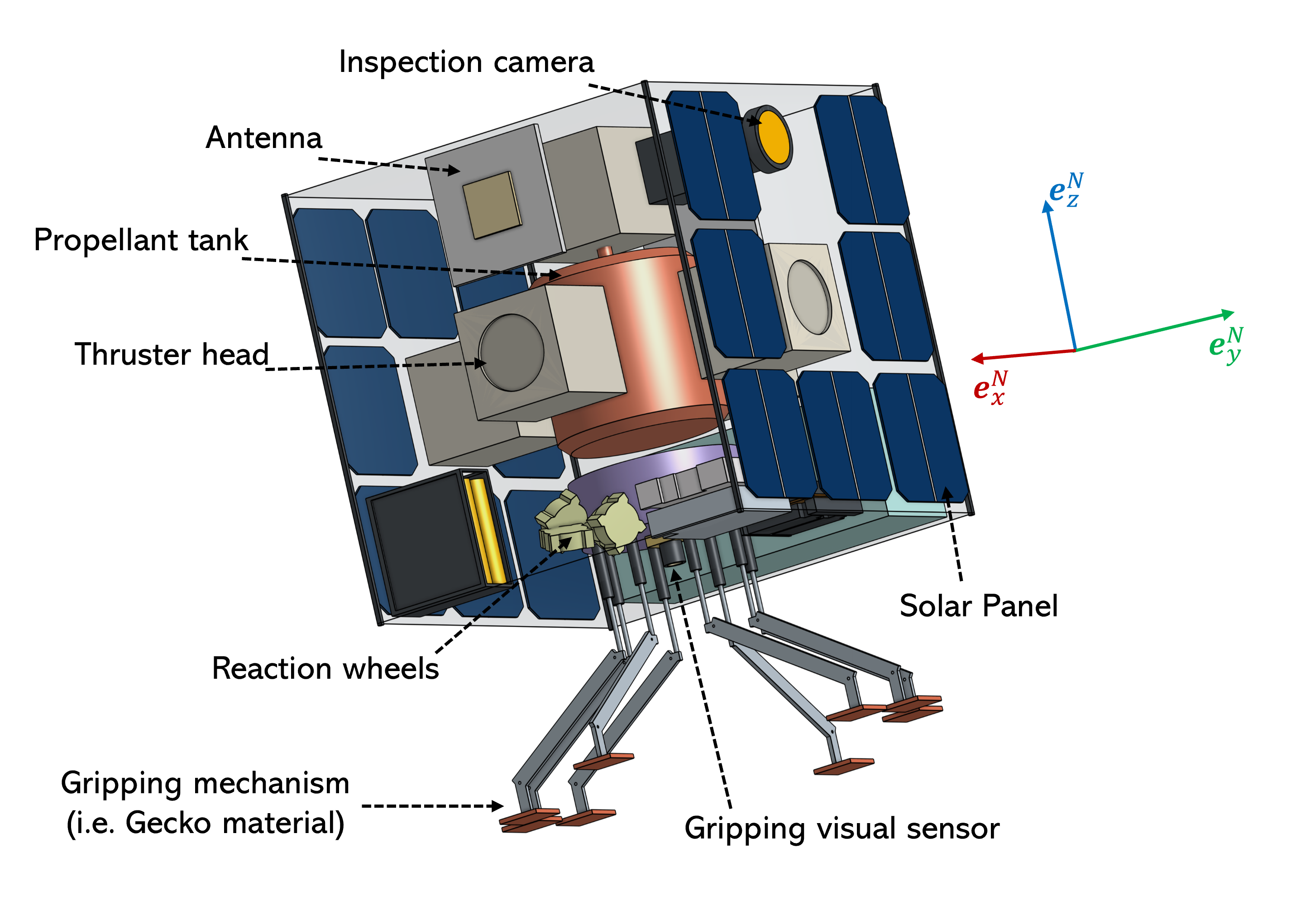}
    \caption{Preliminary 3D Model of the NanoTugs (gripping mechanism inspired by~\cite{hubert2023hybrid}).}
    \label{fig:nanoTugs}
\end{figure}

Table~\ref{tab:subsystems} provides an estimate of the subsystem mass and volume allocation based on a potential selection of components, with a 25\% margin~\cite{LarsonWertz1999}. The sub-total volume and mass of the NanoTugs are approximately 3.5~U and 3.51~kg, respectively. However, the propellant tank will further increase both mass and volume, along with the gripping mechanism and structure, which depend on the final total volume. Additional elements such as wiring, thermal control, and other supporting subsystems will also contribute to the overall budget. The final NanoTug mass and volume are expected to remain within NanoSat/CubeSat limitations~\cite{nasa_cds_rev14_1}. With the given configuration, the dry mass of a NanoTug is estimated to be in the range of $\sim 6$--$10$~kg, accounting for the addition of the gripping mechanism, structural components, and other supporting subsystems to the sub-total mass shown in Table~\ref{tab:subsystems}. This leaves a few kilograms available for propellant mass without exceeding the typical nanosatellites/CubeSat mass range. Figure~\ref{fig:nanoTugs} shows a preliminary 3D model of the proposed NanoTugs, with one thruster head on each of the five faces and the $-Z$ face equipped with the gripping mechanism.

\begin{table}[h!]
\centering
\setlength{\tabcolsep}{4pt} 
\caption{NanoTug Subsystem Volume and Mass Allocation.}
\label{tab:subsystems}
\begin{tabular}{p{8.0cm}cc} 
\Xhline{1pt}
\textbf{Subsystem} & \textbf{Volume (U)} &  \textbf{Mass (kg)} \\
\midrule
On-board computer~\cite{2ndspace2026_master_obc} & 0.22 & 0.16\\
Communications system~\cite{endurosat2026_uhf_transceiver_ii} & 0.25 & 0.12\\
ADCS (reaction wheels)~\cite{cubespace2026cw0057} & 0.15 & 0.60 \\
Electrical power system (battery)~\cite{2ndspace2026_solo_pm4} & 0.50 & 0.60 \\
Inspection camera~\cite{msss2026_ecam_c50} & 0.25 & 0.40 \\
Propulsion system~\cite{DawnAerospace2024Propulsion} & 2.13 & 1.63 \\
\midrule
 \textbf{Sub-Total } & 3.50 & 3.51 \\
\midrule
Others: e.g., propellant, gripping mechanism, structure & - & - \\
\Xhline{1pt}
\end{tabular}
\end{table}

\section{Numerical Study of Swarm-based ADR System}
\label{sec:numerical}
The preliminary design analysis provides a baseline for understanding the NanoTug and swarm sizing for the mission concept. For a more realistic assessment, it is important to consider the dynamics and control of the debris with the NanoTugs distributed on its surface, accounting for the actual thruster placement and propellant consumption. This section presents the NanoTug distribution strategy, the equations of motion for both orbital and attitude dynamics, the control approach with task allocation, and examples that demonstrate the concept connecting it to the preliminary design analysis.

\subsection{Reference Frames and Notations} 
\label{sec:referenceFrames}
\begin{figure}[!h]
    \centering
        \includegraphics[width=0.8\textwidth]{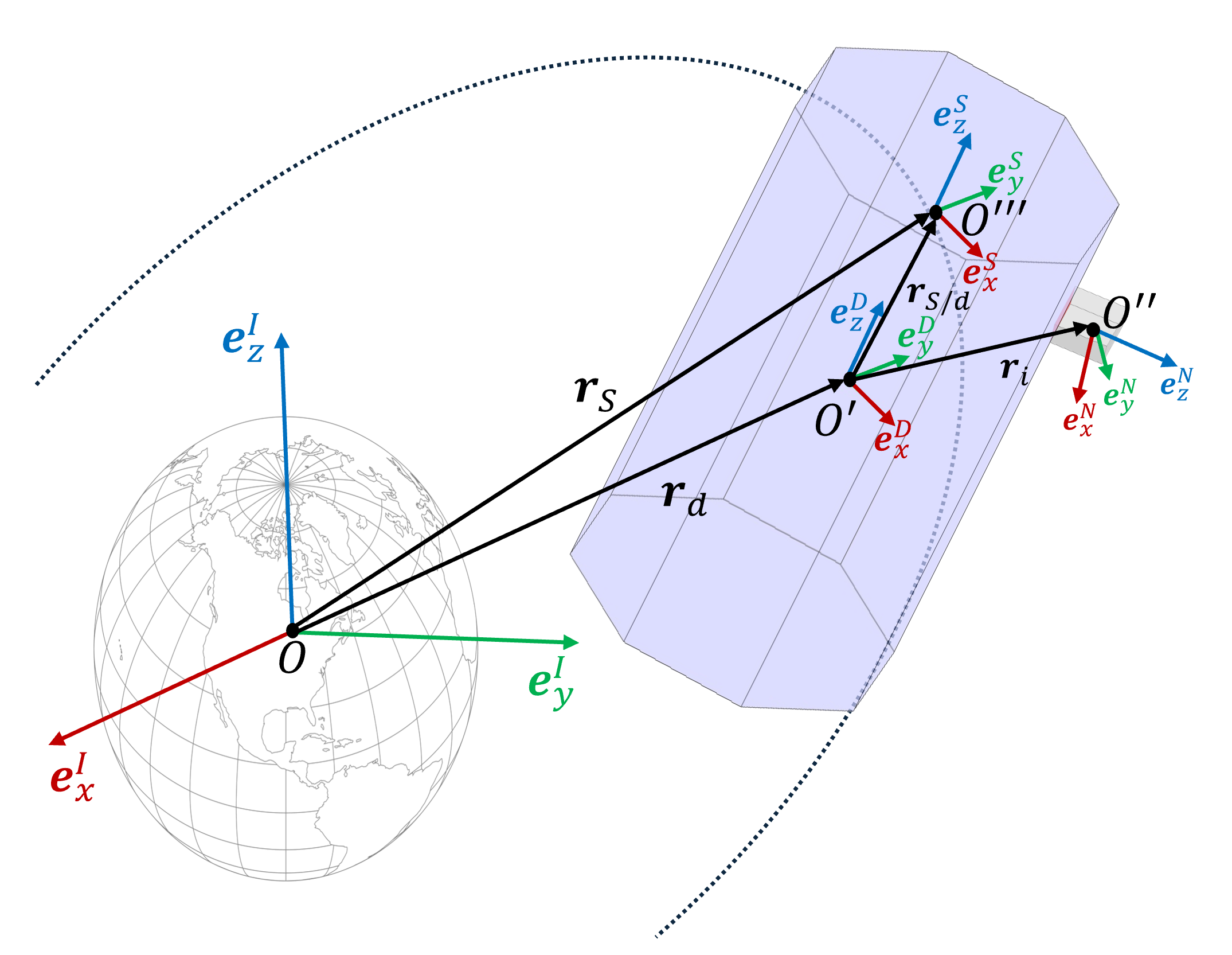}
    \caption{Reference Frames Illustration.}
    \label{fig:refRfames}
\end{figure}

In this work, four different reference frames are used for the stabilization and de-orbiting phase of the proposed mission scenario. Figure~\ref{fig:refRfames} illustrates the reference frames axes and origins.

\begin{enumerate}

    \item The Earth-Centered Inertial (ECI) frame, \(\mathcal{F}^I\): the origin is at the Earth's center of mass $O$, its $\bm e_x^I$-axis is pointed to the Sun at vernal equinox, $\bm e_z^I$-axis is aligned with the Earth's rotation axis, $\bm e_y^I$-axis completes the reference frame to the right-hand triad.

    \item The Debris-Fixed Body frame, $\mathcal{F}^D$, is defined with its origin at the debris's center of mass $O^\prime$, and its axes $\bm{e}_x^D$, $\bm{e}_y^D$, and $\bm{e}_z^D$ are aligned with the principal axes of the debris.
 
    \item The NanoTug-Fixed Body frame, $\mathcal{F}^N$, is defined with its origin at the NanoTug's center of mass $O^{\prime\prime}$, and its axes $\bm{e}_x^N$, $\bm{e}_y^N$, and $\bm{e}_z^N$ are aligned with the principal axes of the NanoTug.
       
    \item The debris-NanoTugs system frame, \(\mathcal{F}^S\): the origin is at the debris-NanoTugs system combined center of mass $O^{\prime\prime\prime}$, and its axes $\bm{e}_x^S$, $\bm{e}_y^S$, and $\bm{e}_z^S$ are aligned with the principal axes of the debris.

\end{enumerate}
The superscripts indicate the reference frame in which a vector is expressed. The subscript $d$ denotes quantities associated with the debris, while the subscript $i$ represents quantities related to the $i^{\text{th}}$ NanoTug. Bold symbols denote vectors, whereas non-bold symbols represent scalar quantities. A hat symbol $\hat{\cdot}$ above a vector indicates that it is a unit vector. $\bm r_d$ is the position vector of the debris center of mass with respect to the origin of the inertial frame, $\bm r_i$ is the position vector of the NanoTug center of mass with respect to the origin of $\mathcal{F}^D$, and $\bm r_s$ is the position vector of the debris-NanoTugs system center of mass with respect to the origin of the inertial frame \(\mathcal{F}^I\). \(\bm r_{S/d}\) denotes the position vector of the center of mass of the debris--NanoTug system with respect to the origin of the debris-fixed frame \(\mathcal{F}^D\). This quantity is required to determine the position of each NanoTug relative to the system center of mass, which is used to compute the torques generated by the NanoTugs about the center of mass for the dynamics and control analysis.
\subsection{NanoTugs Distribution and System Control Capabilities} 
\label{sec:nanoTugsAllocationAndCapabilities}
Once the debris is captured by the NanoTugs, the combined system is treated as a single rigid body with a unified inertia tensor and a combined center of mass, both of which depend on the selected distribution strategy and are used in the dynamics. The system possesses combined thrust and torque capabilities generated by the NanoTugs about system center of mass. This section describes the debris--NanoTugs system properties, the distribution strategies, and the associated control capabilities.

\subsubsection{Debris-NanoTugs System Properties} 
The combined center of mass of the debris--NanoTugs system is
\begin{equation}
\bm{r}^D_{S/d}
=
\frac{
\displaystyle\sum_{i=1}^{N} M_i\,\bm{r}^D_i
}{
M_d + \displaystyle\sum_{i=1}^{N} M_i
}.
\end{equation}
where $M_i = M_{i_\mathrm{dry}} + M_{i_\mathrm{prop}}$ is the wet mass of the NanoTug and $\bm{r}^D_i$ is defined in Fig.~\ref{fig:refRfames}.\\
The debris inertia tensor shifted from the debris center of mass to the combined center of mass is
\begin{equation}
\bm{J}_{d}'
=
\bm{J}_{d}
+
M_d
\left(
(r^D_{S/d})^{2} \mathbf{I}_{3}
-
\mathbf{r}^D_{S/d}\,(\mathbf{r}^D_{S/d})^{\top}
\right),
\end{equation}
\noindent $\bm I_3$ is 3x3 identity matrix.\\ 
The position vector of the $i$th NanoTug center of mass is recalculated as
\begin{equation}
\bm{r'}^D_{i} = \bm{r}^D_i - \bm{r}^D_{S/d},
\end{equation}
and, assuming a point-mass model, its inertia contribution is
\begin{equation}
\bm{J'}_{i} =
M_i
\left((r'^D_{i})^{2} \bm{I}_{3}
-\bm{r'}^D_{i}\,(\bm{r'}^D_{i})^{\top}
\right).
\end{equation}
Therefore, the total inertia tensor about the combined center of mass is
\begin{equation}
\bm{J}
=
\bm{J'}_{d}
+
\sum_{i=1}^{N}\bm{J'}_{i}.
\end{equation}

\subsubsection{NanoTugs Distribution} 

Capturing a tumbling debris object with NanoTugs is a challenging task. After inspection and detailed characterization of the target, the NanoTugs can identify optimal attachment points to achieve effective attitude and orbital control. However, accurately gripping these designated points is often challenging and depends on several factors, including the debris surface material and geometry, the debris’ tumbling motion, and other operational constraints.

\begin{figure}[!h]
    \centering
    \includegraphics[width=0.8\textwidth]{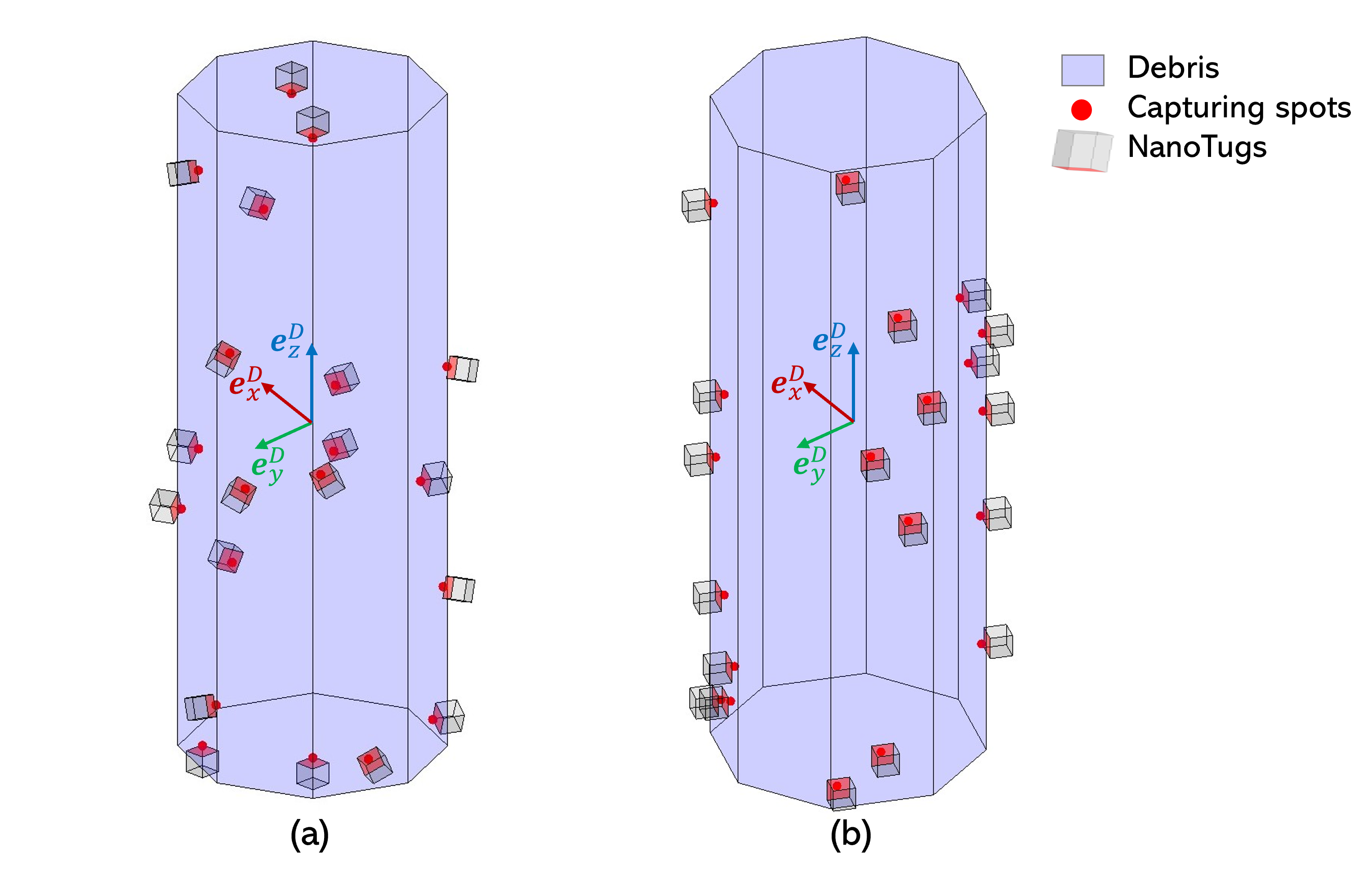}
    \caption{Debris with (a) Random and (b) Predefined Allocation of 20 NanoTugs.}
    \label{fig:DebrisnanoTugs}
\end{figure}

The distribution of the NanoTugs on the debris surface is important and directly affects the control capabilities of the swarm. Different distribution strategies can be considered depending on the control approach. In this study, two strategies for NanoTug distribution on the debris surface are considered:

\begin{enumerate}
    \item \textit{Random distribution:} In this case, we explore the idea of uniformly distributing the NanoTugs across the debris surface. Since the capturing process is inherently imprecise in terms of gripping location, we ideally assume that the same number of NanoTugs is assigned to each debris face, which is represented by an octagonal prism shape (a better representation can be achieved, e.g., using a hexadecagonal prism to more closely simulate the cylindrical shape of upper-stage rocket bodies). The particular location of a NanoTug on each face is random, following a uniform distribution over that face. If the total number of NanoTugs cannot be evenly divided among the faces, the remaining NanoTugs are assigned following the indexing of the faces. The random distribution is considered because targeting specific capture points can be challenging, especially with tumbling debris, a topic that can be studied in more detail later.

    \item \textit{Predefined distribution:} NanoTugs are distributed across selected debris faces such that each NanoTug has one thruster aligned with the desired de-orbiting attitude of the debris. This attitude is targeted by the control law so that the aligned thrusters are pointed approximately in the anti-velocity direction, maximizing the reduction in the orbit's semi-major axis. The remaining thrusters on each NanoTug are used for attitude control, including de-tumbling and to ensure that the debris maintains the desired attitude throughout the de-orbit maneuver.
\end{enumerate}
Figure~\ref{fig:DebrisnanoTugs} illustrates the (a) uniform random distribution of $N$ NanoTugs across the debris surface and (b) the predefined distribution of the NanoTugs. In Fig.~\ref{fig:DebrisnanoTugs} (b), the showed distribution allows all NanoTugs to produce thrust along $-\bm{e}_x^D$. During de-orbiting, the debris can be oriented such that $\bm{e}_x^D$ points in the velocity direction, maximizing the efficiency of the orbit-lowering maneuver. In both distribution strategies, the NanoTugs are maintained at a minimum separation distance and ensuring that the debris surface area is sufficient to accommodate the required number of NanoTugs.

\subsubsection{NanoTugs Control capabilities} 

As shown in the 3D model in Fig.~\ref{fig:nanoTugs}, each NanoTug is equipped with five thruster heads, with the corresponding unit thrust vectors \(\bm{T}_{ij}^N\) defined in the NanoTug body frame \(\mathcal{F}^N\), where \( j = 1, \ldots, 5 \) denotes the thruster index for the \( i^{\text{th}} \) NanoTug:
\begin{equation}
\begin{bmatrix} 
\hat{\bm T}_{i1}^N & \hat{\bm T}_{i2}^N& \hat{\bm T}_{i3}^N & \hat{\bm T}_{i4}^N & \hat{\bm T}_{i5}^N 
\end{bmatrix} =  
\begin{bmatrix}
1 & 0 & 0 \\
-1 & 0 & 0 \\
0 & 1 & 0 \\
0 & -1 & 0 \\
0 & 0 & 1 \\
\end{bmatrix}^\top
\end{equation}
The transformations of the vectors between frames are done by unit quaternion, where quaternion is defined as follows:
 \begin{equation}
\bm q = q_0 + \bar{\bm q}
\end{equation}
\noindent $q_0$ is the scalar part and $\bar{\bm q}$ is the three-dimensional vector part of the quaternion $\bm q$.

The thrust vector of each head is transformed to the debris frame $\mathcal{F}^D$ as follows:

\begin{equation}
\bm{T}_{ij}^D = \bm{q}_{DN} \circ \bm{T}_{ij}^N\circ \tilde{\bm{q}}_{DN},
\label{eq:thrustNanoTugsD}
\end{equation}
\noindent where $\bm{T}_{ij}^D$ and $\bm{T}_{ij}^N$ are the vector parts of the quaternions with zero scalar part, "$\circ$" denotes quaternion product, $\bm{q}_{DN}$ is the unit quaternion representing the rotation from the NanoTug body frame $\mathcal{F}^N$ to the debris body frame $\mathcal{F}^D$ and $\tilde{\bm{q}}_{DN}$ is the quaternion conjugate. $\bm{q}_{DN}$ is defined such that the NanoTug body $z$-axis, $\bm e_z^N$, along which the gripping mechanism is oriented in the negative direction, is perfectly aligned with the surface normal of the debris at the designated gripping point.






Similarly, the Thrust vectors can be transformed to $\mathcal{F}^I$ as follows:

\begin{equation}
\bm{T}_{ij}^I = \bm q_d \circ \bm{T}_{ij}^D \circ \tilde{\bm{q}}_d,
\label{eq:thustInInertial}
\end{equation}
\noindent where $\bm{T}_{ij}^I$ and $\bm{T}_{ij}^D$ are the vector parts of the quaternions with zero scalar part, $\bm q_d$ is the unit quaternion of the debris with respect to $\mathcal{F}^I$.

The combination of all the NanoTugs thruster heads thrust vectors attached to the debris forms a 3D matrix, which can be expressed in any desired reference frame.

\begin{equation}
\mathcal{T}_{kji} = \vcenter{\hbox{\includegraphics[height=4cm]{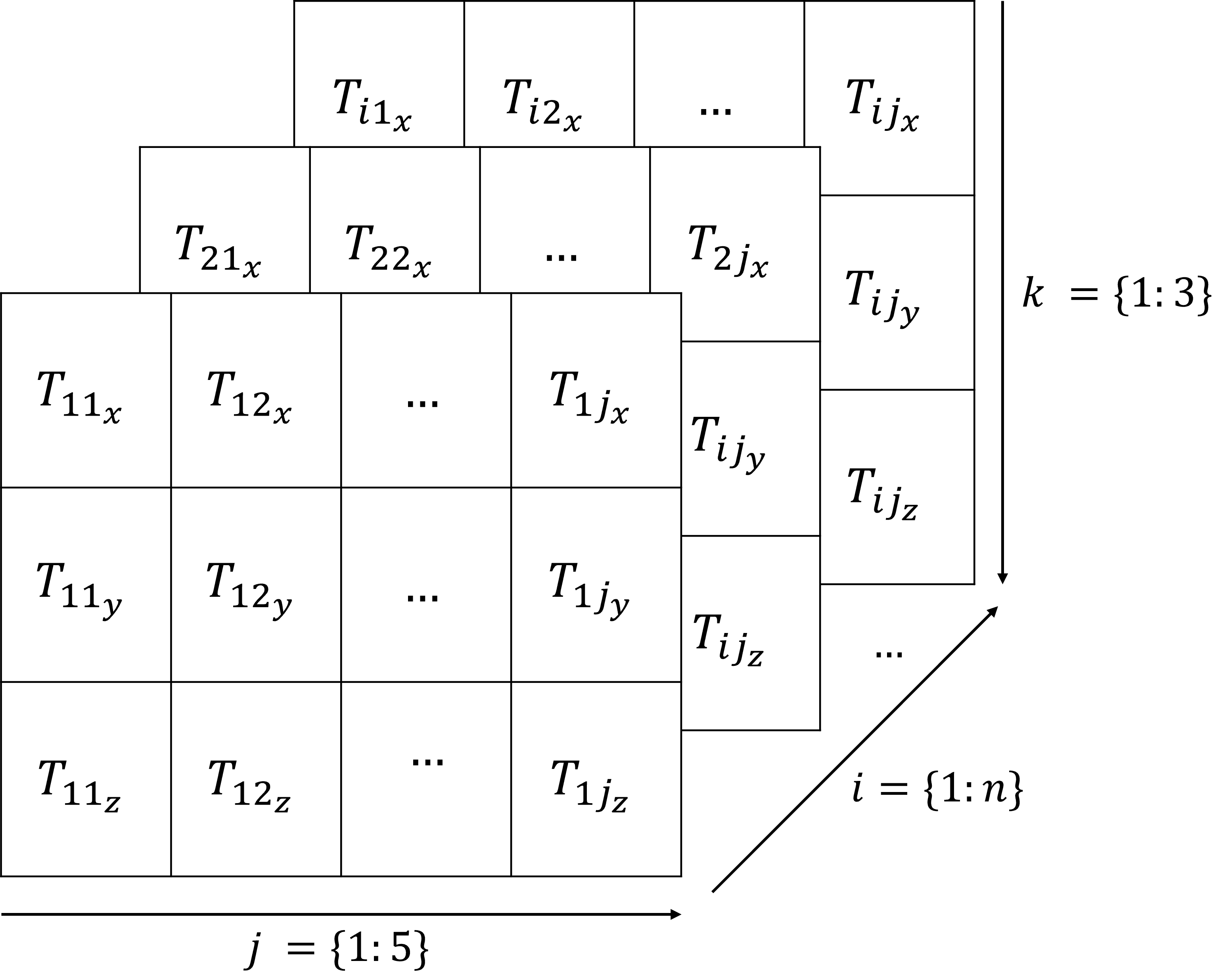}}}
\end{equation}

\noindent where $k = 1,2,3$ denotes the $x$, $y$, and $z$ components of the vector $\bm T_{ij}$, $j = 1,\ldots,5$ is the thruster head index of a NanoTug, and $i = 1,\ldots,n$ is the index of the NanoTug attached to the debris.\\

The torque generated by each thruster head in the debris frame is computed as:
\begin{equation}
\bm M_{ij}^D = \bm{r}^D_{i} \times \bm T_{ij}^D
\end{equation}
Similarly, the combination of the NanoTugs torque vectors forms a 3D matrix:

\begin{equation}
\mathcal{M}_{kji} = \vcenter{\hbox{\includegraphics[height=4cm]{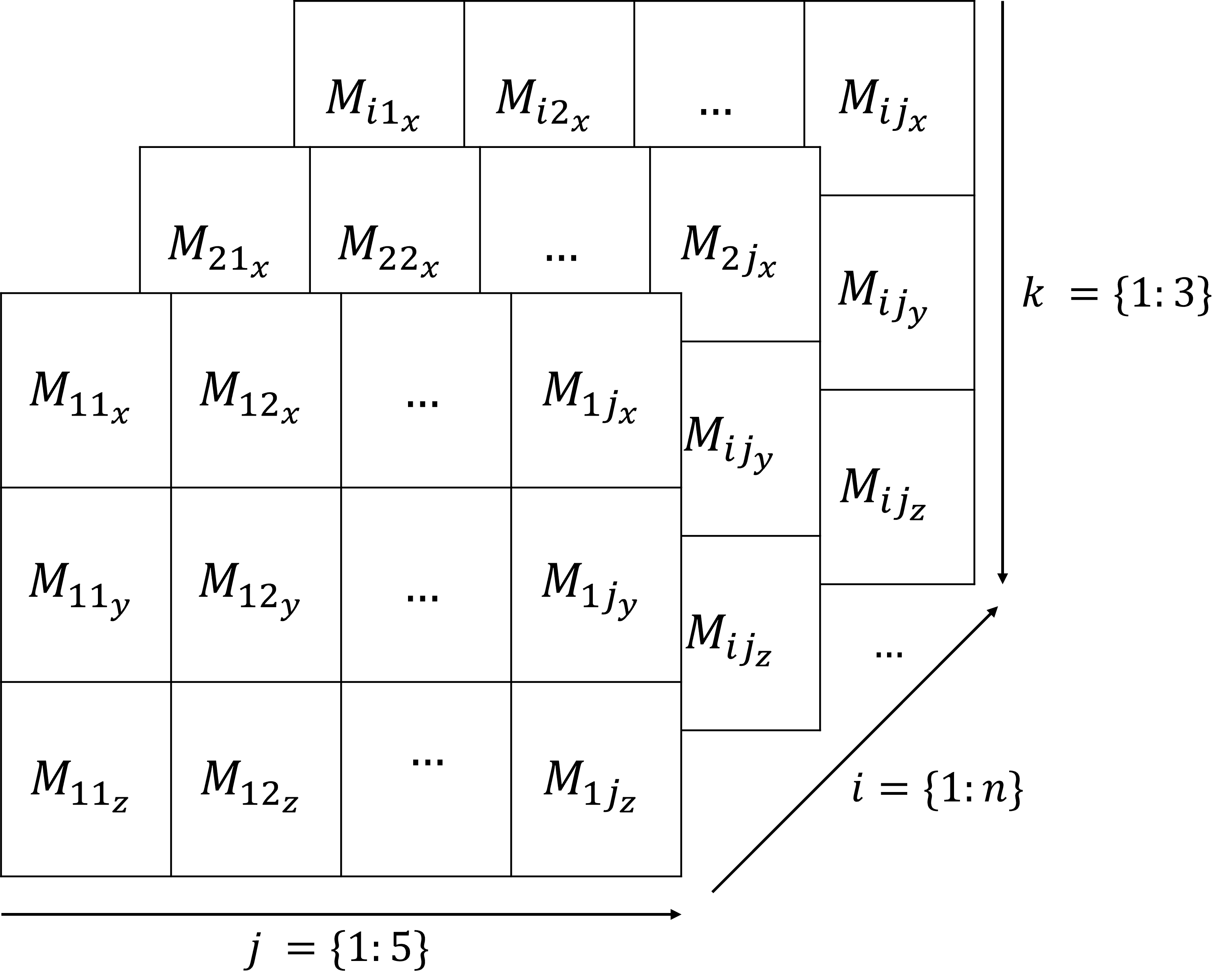}}}
\end{equation}
It should be noted that in attitude dynamics the torque should be provided with respect to the system center of mass frame $\mathcal{F}^S$, however, it is assumed that with the given NanoTugs distribution, the shift in center of mass is not significant so $\mathbf{r'}^D_{i} \approx  \mathbf{r}^D_{i}$ and $\bm M_{ij}^S \approx  \bm M_{ij}^D$

\subsection{Equations of Motion} 
\label{sec:EquationsOfMotion}
In this study, the debris and attached NanoTugs are modeled as a single integrated system for both orbital and attitude dynamics. The translational and rotational motions are coupled, as thrust from the NanoTugs influences both the system center of mass motion and the attitude. The dynamics account for the combined inertia tensor of the debris–NanoTug system, considering the contributions of each NanoTug. Changes in the system's mass properties—including total mass and inertia tensor—due to propellant consumption and NanoTug placement are also considered throughout the de-orbit maneuver.
\subsubsection{Orbital Dynamics}
The following equation describing orbital motion dynamics in the ECI frame $\mathcal{F}^I$ is used:

\begin{equation} \label{eq:equation of motion}
    \ddot{\mathbf{R}} = - \frac{\mu \mathbf{R}}{R^3} + \mathbf{a}_{drag} + \mathbf{a}_{J_2} + \mathbf{a}_c,
\end{equation}

\noindent $\mathbf{R}$ is the position vector of the debris-NanoTugs center of mass with respect to the inertial frame, $\mathbf{a}_{drag}$ is the atmospheric drag acceleration, $\mathbf{a}_{J_2}$ stands for external disturbance caused by the Earth oblateness and $\mathbf{a}_c$ is the control acceleration \cite{curtis2013orbital, alfriend2009spacecraft}.

\begin{equation}\label{eq:drag_perturbation}
\mathbf{a}_{drag} = -\frac{1}{2} \ \rho \ v \left( \frac{C_d \ A}{M_{S}} \right) \bm v,
\end{equation}

\noindent where $\rho$ is the atmospheric density, $C_d$ is the debris dimensionless drag coefficient, $M_S$ is the total mass of the system including both the debris and the NanoTugs which changes based on Eq.~\ref{eq:mass as time}, and $\bm v = \mathbf{V} - \boldsymbol{\omega_{\oplus}}\times \mathbf{R}$ is the velocity of the debris relative to the atmosphere, $\boldsymbol{\omega_{\oplus}}$ is the Earth self-rotation angular velocity. In this work, a constant drag coefficient $C_d = 2.2$ is assumed for all debris surfaces~\cite{LarsonWertz1999}. The drag area $A$ represents the effective projected area normal to $\bm v$ and is computed by summing the projected contributions of all surfaces of the debris geometry, modeled as an octagonal prism, such that
\begin{equation}
A = \sum_i (\hat{\mathbf{N}}_{d_i} \cdot \hat{\bm v}) \, A_{d_i} ,
\end{equation}
\noindent where $A_{d_i}$ is the area of the $i$-th debris facet, $\hat{\mathbf{N}}_{d_i}$ is its unit facet normal vector, and $\hat{\bm v}$ is the unit vector in the direction of the relative velocity. Only surfaces with positive projection $(\hat{\mathbf{N}}_{d_i} \cdot \hat{\bm v} > 0)$ are considered. Atmospheric density is computed using the piece-wise exponential model from~\cite{vallado2001fundamentals}. The drag due to NanoTugs is not considered in this study.

\begin{equation}\label{eq:J2_perturbation}
\mathbf{a}_{J_2} = \frac{3\mu J_2 \mathcal{R}^2_{\oplus}}{2 R^5}\left[ \left( \frac{5Z^2}{R^2} -1 \right)\mathbf{R} - 2\mathbf{Z} \right],
\end{equation}

\noindent where $J_2 = 1.08263 \times10^{-3}$ is the second zonal harmonic of Earth, $\mathcal{R}_{\oplus} = 6378.137 \ \text{km}$ is the mean Earth equatorial radius  and $\mathbf{Z} = [0, 0, Z]^{\top}$. 

It should be noted that the other disturbing forces-induced accelerations caused by solar radiation pressure, and third bodies gravitation are not considered yet in this study. This can be considered in future work for a more accurate analysis.

\subsubsection{Attitude Dynamics}

The rigid body quaternion kinematical equations and rigid body dynamics equations used in the stabilization and de-orbiting phase for propagating the change in rigid body orientation and angular velocity are as follows:
\vspace{10pt}
\begin{equation}
\dot{\bm q} = \frac{1}{2} \ \bm q \ \circ 
\bm{\omega}
\end{equation}

\noindent where $\bm{\omega}$ is quaternion vector part of rigid body angular velocity in body fixed reference frame with zero scalar part.
\vspace{10pt}
\begin{equation}
\dot{\bm{\omega}} = \bm{J}^{-1} \left( \bm{\tau} + \bm L - \bm{\omega} \times \bm{J} \bm{\omega} \right)
\label{eq:wdot}
\end{equation}
\noindent where $\bm J$ is the system inertia tensor, $\bm{\tau}$ is the active control system torque, and $\bm L$ represents the environmental disturbing torques, both expressed in the body-fixed frame.

Based on the multi-mission environmental torque analysis methodology presented in \cite{Alnaqbi2022IAC}, the dominant sources of attitude disturbance torques for cylindrical debris, such as rocket stages, in low Earth orbit (LEO) are aerodynamic and gravity-gradient torques. Accordingly, only these two disturbance sources are considered in the present study, as they represent the largest contributors to the attitude disturbances for this type of debris.

The debris body is discretized into a number of elements, where the forces and torques acting on each element are computed, and the total disturbance torque is obtained by summing all individual contributions. In this study, the \emph{mass model} of the debris which consists of mass elements is employed to compute torques arising from gravity--gradient effects, while the \emph{surface model} of the debris which consists of surface elements, is used to calculate the torques due to aerodynamic forces. The effect of NanoTugs on the disturbance torques is not considered this work.

The aerodynamic force $d\bm F_{\text{Aero}}$ is calculated on each surface element with area $A_{s}$ with surface normal $\hat{\mathbf{N}_s}$ as follows~\cite{wertz_adcs_1978}:

\begin{equation}
d\bm F_{\text{Aero}} = -\frac{1}{2} C_d \rho \ \bm v_{s}^2 (\hat{\mathbf{N}_s} \cdot \hat{\bm v}_{s})\,\hat{\bm v}_{s}\, A_{s}
\label{eq:dFaero}
\end{equation}

\noindent where $\hat{\bm v}_{s}$ is the unit vector in the direction of the translational velocity $\bm v_{s}$ of the surface element relative to the incident stream, $\rho$ is the atmospheric density, and $C_d$ is the drag coefficient.
The term $(\hat{\mathbf{N}_s} \cdot \hat{\bm v}_{s})$ accounts for the projection of the surface element area $A_{s}$ onto the direction of the incoming flow.
For a spinning debris with angular velocity $\bm \omega$, the translational velocity of the surface element is:
\begin{equation}
\bm v_s = \bm v + \bm \omega \times \bm r_s
\label{eq:vTrans}
\end{equation}

\noindent where $\bm v$ is the velocity of the debris center of mass relative to the atmosphere and $\bm r_s $ is the vector from debris center of mass to the surface element.

The aerodynamic torque $\bm L_{\text{Aero}}$ acting on the debris is as follows, where the torque considered only for the surface elements with $(\hat{\mathbf{N}_s} \cdot \hat{\bm v}_{s}) > 0$:

\begin{equation}
\bm L_{\text{Aero}} = \int \bm r_s \times d\bm F_{\text{Aero}}
\label{eq:Naero}
\end{equation}

The gravitational force $d\bm F_{\text{GG}}$ acting on mass element with mass $M_m$ is as follow~\cite{wertz_adcs_1978}:

\begin{equation}
d\bm F_{\text{GG}} = -\frac{\mu \bm R_m }{R_m^3}M_m
\label{eq:dFGravGrad}
\end{equation}

\noindent where $\bm R_m$ is the vector from Earth center to the mass element. 
The total gravity gradient torque acting on the debris is:
\begin{equation}
\bm L_{\text{GG}} = \int \bm r_m \times d\bm F_{\text{GG}}
\label{eq:NGG}
\end{equation}

\noindent where $\bm r_m$ is the vector from the geometric center of the debris to the mass element in the debris \emph{mass model}.

\subsection{Debris Orbital and Attitude Control Algorithms} 
\label{sec:debrisControl}
Once the NanoTugs are distributed across the debris surface, the stabilization and de-orbiting phase is initiated. The first objective in this stage is to reduce the debris’ angular velocity, making it easier to lower its altitude using the thrusters. During de-orbiting, at each control loop, on/off flags are assigned to each thruster on each NanoTug to activate only those generating thrust in the required direction for de-orbiting. Simultaneously, if attitude control is required, the thrusters generating torque in the desired direction are activated to prevent excessive angular velocity or maintain the required attitude if needed. Due to power limitations, only one thruster per NanoTug can be activated at a time. This represents a preliminary control approach used to demonstrate the concept; more advanced control logic should be investigated in future studies.

The required torque either to stabilize the debris or point to certain attitude $\bm q_r$ is calculated a follows 

\begin{equation}
    \bm u_{us} = -K \delta\bm{\overline{q}} - \bm P \delta\bm\omega 
    + \bm J\left(\dot{\bm\omega}_r - \tilde{\bm\omega}\bm\omega_r\right) 
    + \tilde{\bm\omega}\bm J\bm\omega - \bm L
    \label{eq:unsat_torque}
\end{equation}

\noindent where $\bm u_{us}$ is the unsaturated control torque in debris body frame, $K$ is the attitude error feedback gain, $\delta\bm{\overline{q}}$ is the vector part of the quaternion attitude error $\delta \bm q = \tilde{\bm{q}}_{r} \circ \bm{q}$, $\bm P$ is the damping gain matrix, $\delta \bm\omega = \bm\omega - \bm\omega_r$ is the angular velocity tracking error, $\bm J$ is the system inertia tensor, $\bm\omega_r$ and $\dot{\bm\omega}_r$ represent the reference angular velocity and its derivative, $\tilde{\bm\omega}$ is the skew-symmetric matrix associated with the angular velocity $\bm\omega$, and $\bm L$ is any other external disturbance torques ~\cite{schaub2003analytical}.

As discussed in Section~\ref{sec:nanoTugsAllocationAndCapabilities}, two NanoTug distribution strategies are considered: random distribution and predefined distribution. For the random distribution, the NanoTugs produce thrust in arbitrary directions. In this case, identifying a targeted debris attitude during stabilization to maximize de-orbit efficiency is not practical. Many NanoTugs will not contribute effectively, as their thrusters may not be aligned with the anti-velocity direction when the targeted attitude is achieved. This is especially true if the debris’ angular velocity has been reduced to near zero; only a small subset of randomly distributed NanoTugs will actively contribute to the de-orbit maneuver. Therefore, no specific attitude control is considered during stabilization or de-orbiting, and the attitude error feedback gain $K$ is set to zero. In the other hand, for the predefined distribution, the NanoTugs are carefully positioned so that their thrusters align with the anti-velocity vector when the debris is oriented to a targeted attitude. In this scenario, the control system actively stabilizes the debris to the desired attitude to try to increase de-orbit efficiency. A large percentage of NanoTugs contribute effectively during the maneuver, and the targeted attitude is continuously maintained throughout the de-orbiting process. In both cases, $\bm \omega_r = [0, 0, 0]^\top$ and  $\dot{\bm \omega}_r = [0, 0, 0]^\top$. Under these conditions, Eq.~\ref{eq:unsat_torque} simplifies to:

\begin{equation}
\bm u_{us} = -K \delta\bm{\overline{q}} -\bm P\delta\bm\omega + \tilde{\bm\omega}\bm J \bm\omega - \bm L 
\label{eq:unsat_torqueDet}
\end{equation}
Based on the required control torque derived from Eq.~\ref{eq:unsat_torqueDet} and the NanoTugs distribution, an on/off flag \(f_{M_{ij}}\) is assigned to each thruster head of every NanoTug. The selection criterion depends on the alignment between the torque vector produced by each head, \(\bm M_{ij}^D\), and the required control torque vector, \(\bm u_{us}\), as follows:

\begin{equation}
    f_{M_{ij}} =
    \begin{cases} 
    1 \ \text{(on)}, & \text{if } \hat{\bm M}^D_{ij} \cdot \hat{\bm u}_{us} \geq a_{M_{\min}} \ \& \ M_{i_{\mathrm{prop}}} > 0 \\[1mm]
    0 \ \text{(off)}, &  \text{otherwise}
    \end{cases}
\label{eq:nozzleFlagTorque}
\end{equation}

\noindent where \( a_{M_{\min}} = \cos(\theta_{M_{\min}}) \) represents the minimum alignment threshold between the desired torque direction and each thruster head torque. The angle \( \theta_{M_{\min}} \) should be selected to avoid inefficient torque contributions. It is assumed that only one thruster per NanoTug can be active at each control step. If more than one thruster satisfies the condition within the same NanoTug, only the one with the highest alignment with the required torque direction is selected, and the others are turned off.

\noindent Based on the selected thrusters, the total achievable torque, denoted as \( \bm u_{\mathrm{avail}} \), is computed as
\begin{equation}
\bm u_{\mathrm{avail}} = (u_1, u_2, u_3)^\top,
\label{eq:availableTorque}
\end{equation}
\[
u_k = \mathcal{M}^D_{kij} \, f_{ij}.
\]
The available torque \( \bm u_{\mathrm{avail}} \) computed from Eq.~\ref{eq:availableTorque} may be either smaller or larger than the required control torque \( \bm u_{us} \). If the available torque is smaller than the required torque for at least one component \( k \), the commanded torque is limited by the available torque:
\[
\bm \tau = \bm u_{\mathrm{avail}}, \quad \text{if any } |u_{us,k}| > |u_{\mathrm{avail},k}|.
\]
If the available torque exceeds the required torque, the active thrusters are ranked based on their projection onto the desired torque direction. Thrusters with the highest projections are retained, and others are deactivated. The torque is recomputed by summing the contributions of the selected thrusters. This process continues iteratively until all components of the resulting torque \( \bm u_{\mathrm{avail}}'\) match or just exceeds the required torque. The final commanded torque \( \bm \tau \) is 
\[
\bm \tau = \bm u_{\mathrm{avail}}', \quad \text{if all }|u_{\mathrm{avail},k}| > |u_{us,k}|. 
\]
It should be noted that the available torque will not, in general, be equal to the required torque. This is due to the alignment threshold defined in Eq.~\ref{eq:nozzleFlagTorque}, whereby activated thrusters may also generate torque components in other directions. In addition, the thrusters considered in this work are non-throttleable and operate only at their maximum thrust level, which further limits the ability to precisely match the required torque.

On the other hand, for de-orbiting, the Gauss' variational equations gives the following expression for the change in semi-major axis $a$:
\begin{equation}
\frac{da}{dt} = \frac{2a^2 v}{\mu} a_v,
\label{eq:changeinSMA}
\end{equation}

\noindent where $a_v$ is the projection of the control acceleration in the velocity direction. Based on Eq.~\ref{eq:changeinSMA}, to achieve maximum change in semi-major axis, the projection of the control acceleration should be maximized in the orbit velocity direction $a_v$. 
In this case, the ``on'' flag is assigned to the thruster heads whose thrust vectors have a component directed opposite to the debris velocity vector.
\begin{equation}
    f_{T_{ij}} =
    \begin{cases} 
    1 \ \text{(on)}, & \text{if }  \hat{\bm T}_{ij}^I \cdot -\hat{\bm V}_{d}^I \geq a_{T_{min}} \  \& \  \  M_{i_{\mathrm{prop}}} > 0 \\[1mm]
    0 \ \text{(off)}, & \text{otherwise}
    \end{cases}
\label{eq:nozzleFlagThrust}
\end{equation}

\noindent where \(a_{T_{min}} = \cos(\theta_{T_{min}})\) represents the minimum alignment threshold between $-\hat{\bm V}_{d}^I$ and the thrust vector of each thruster in \(\mathcal{F}^I\). Similarly, the threshold angle $\theta_{T_{\min}}$ is chosen below $45^\circ$ so that at most one thruster head can satisfy the selection condition, while minimizing deviations from the required thrust direction. If $\theta_{T_{\min}} \geq 45^\circ$, the thruster head with the highest projection within a NanoTug is used only.

Given the flags obtained from the attitude control, \( f_{M_{ij}} \), and the orbit control, \( f_{T_{ij}} \), the final flag matrix is defined by combining both while ensuring that at most one thruster per NanoTug is active
\begin{equation}
    f_{ij} = f_{M_{ij}} + f_{T_{ij}}.
\label{eq:nozzleFlagTot}
\end{equation}
Based on the resulting flag matrix, the control torque \( \bm \tau \) and control acceleration \( \bm a_c \) are computed as follows:

\begin{equation}
\bm \tau = (\tau_1, \tau_2, \tau_3)^\top,
\label{eq:inputtorqueDeOrbiting}
\end{equation}
\[
\tau_k = \mathcal{M}^D_{kij} \, f_{ij}.
\]

\begin{equation}
\bm a_c = (a_1, a_2, a_3)^\top,
\label{eq:inputAcceleration}
\end{equation}
\[
a_k = \frac{\mathcal{T}^I_{kij} \, f_{ij}}{M_S}.
\]
\subsection{Simulations and Results}
\label{sec:simResults}
\begin{figure}[!h]
    \centering
    \includegraphics[width= 0.8\textwidth]{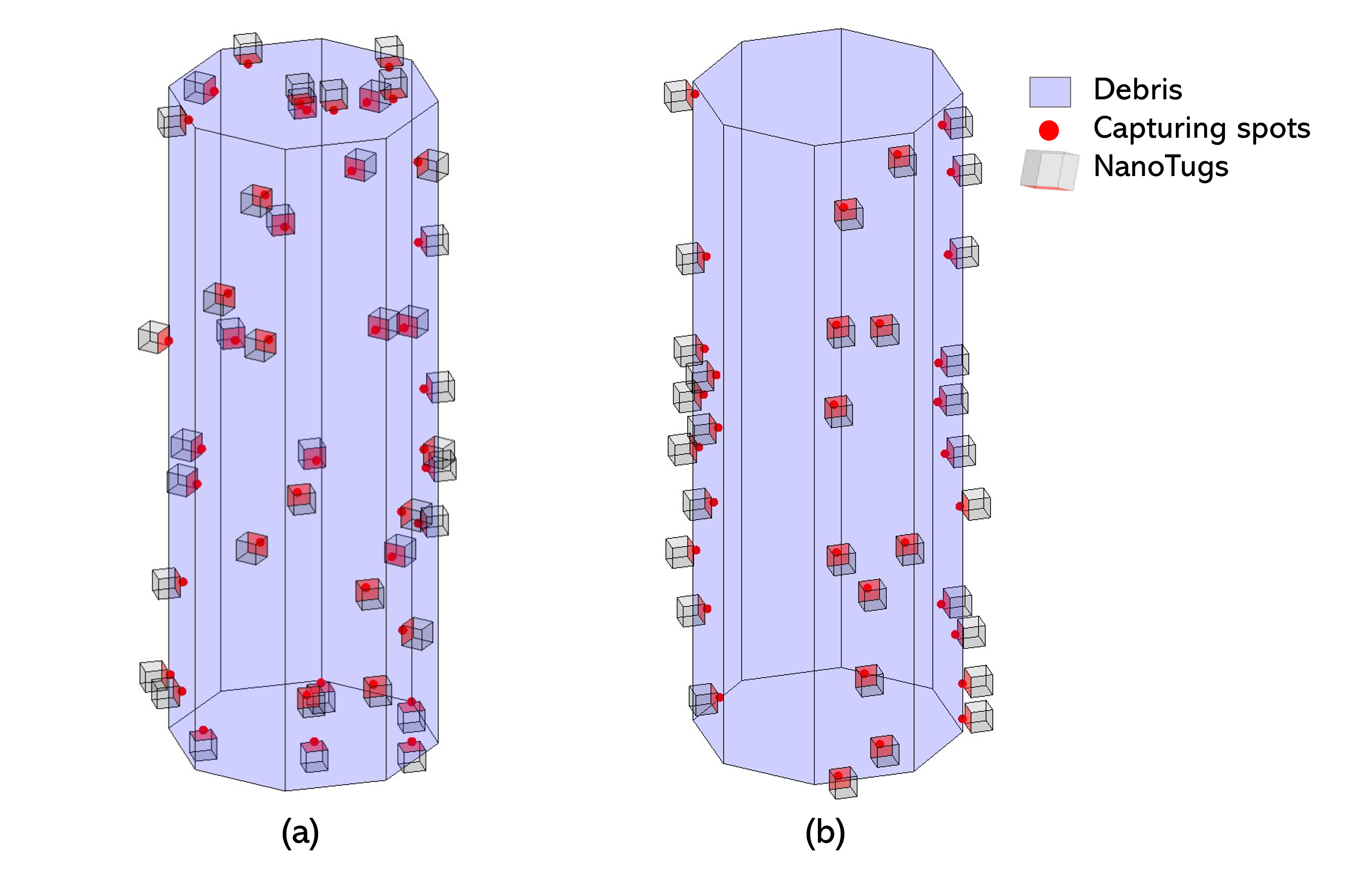}
    \caption{NanoTugs Distribution (a) Random, (b) Predefined.}
    \label{fig:nanotugsDist}
\end{figure}

This section presents an example of applying the analytical system sizing and mission evaluation approach described in Section~\ref{sec:conceptOverview}, and connects it to the numerical simulation methodology outlined in Section~\ref{sec:numerical}. The SL-8 R/B debris object~\cite{eusst2021_sl8rb_ops6182, satcat2025_sl8rb_9023} is selected as a case study, using its orbit, mass, and dimensions. The analysis is structured by first defining mission and candidate system parameters, followed by the calculation of required propellant, verification of feasibility, and determination of the number of NanoTugs. Table~\ref{tab:systemSizing} summarizes the key input parameters and the resulting system sizing calculations for this example following the system sizing steps. Two cases with different $\lambda$ are considered. The first case which is $\lambda = 0.8$ is considered for the random NanoTugs distribution scenario and the second case $\lambda = 1$ is considered for the predefined NanoTugs distribution. Figure~\ref{fig:nanotugsDist} shows the distribution of the NanoTugs across the debris surface for both cases.\\

\begin{table}[h!]
\centering
\caption{System sizing example for SL-8 R/B debris for two $\lambda$ cases.}
\label{tab:systemSizing}
\begin{tabular}{l l c c c}
\Xhline{1pt}
\textbf{Category} & \textbf{Parameter} & \textbf{Case 1} & \textbf{Case 2} & \textbf{Unit} \\
 &  & ($\lambda = 0.8$) & ($\lambda = 1$) & \\
\hline
\multirow{7}{*}{Mission \& Debris} 
 & $M_d$ & \multicolumn{2}{c}{1434} & kg \\
 & $h_0$ & \multicolumn{2}{c}{760} & km \\
 & $h_f$ & \multicolumn{2}{c}{300} & km \\
 & $t_\text{expected}$ & \multicolumn{2}{c}{8} & hours \\
 & $H_d$ & \multicolumn{2}{c}{6.3} & m \\
 & $R_d$ & \multicolumn{2}{c}{1.2} & m \\
 & $\mathbf{J}_d$ & \multicolumn{2}{c}{$\operatorname{diag}(5259.2,\, 5259.2,\, 1032.5)$} & kg m$^2$ \\
\hline
\multirow{4}{*}{Candidate System} 
 & $M_{i_\mathrm{dry}}$ & \multicolumn{2}{c}{6} & kg \\
 & $I_\mathrm{sp}$ & \multicolumn{2}{c}{200} & s \\
 & $T_i$ & \multicolumn{2}{c}{0.5} & N \\
 & $\lambda$ & 0.8 & 1 & - \\
\hline
\multirow{4}{*}{Calculated / Derived} 
 & $t_\mathrm{control}^{\max}$ & 8 & 8 & hours \\
 & $M_{i_\mathrm{prop,0}}$ & 5.8736 & 7.3420 & kg \\
 & $n$ & 31 & 30 & - \\
 & $N$ & 39 & 30 & - \\
\Xhline{1pt}
\end{tabular}
\end{table}

\noindent\textit{Random NanoTugs Distribution} \\
The first scenario considers the concept of a random NanoTugs distribution discussed earlier in this section. From the analytical approach for the system sizing, using the mission parameters and the selected NanoTugs system in Table~\ref{tab:systemSizing} with $\lambda = 0.8$, it is shown that in order to de-orbit the debris from $h_0$ to $h_f$ within 8~hours, an average minimum of 31 NanoTugs must operate simultaneously, while the total number of NanoTugs is 39. Each NanoTug should carry at least $5.8736$~kg of propellant. The selection of $\lambda$ was based on several numerical simulation observations with the random distribution.

Using these parameters, the numerical simulation was carried out, taking into account the disturbance torques from aerodynamic forces and gravity gradients. Table~\ref{tab:simParametersRandom} shows simulation parameters for this scenario. The initial angular velocity of the debris is set to $20$~deg/s along each axis. To account for assumptions made in the analytical system sizing, $10\%$ is added to both the total number of NanoTugs and the propellant carried by each NanoTug. This is done to take into account that the thrusters are not be perfectly aligned with the anti-velocity direction, that not all $n$ NanoTugs will be active at each control step, and that some propellant will be consumed for attitude maintenance and to counter external disturbance torques. All of these effects were neglected in the analytical system sizing.

A maximum allowable angular velocity error of $5$~deg/s is set to ensure that all NanoTugs can contribute to the de-orbiting mission and to not spend excessive propellant trying to stabilize the debris to a very low rotation rate, since the primary mission goal is de-orbiting. The control gain $K$ is set to zero due to the random distribution, as discussed earlier. In this scenario, the attitude control torque is only applied if
\[
\left| \bm \omega_{d,i} - \bm \omega_{r,i} \right| > \delta\omega_{\mathrm{max}}, \quad \forall i \in \{x, y, z\}.
\]

\begin{figure}[!h]
    \centering
    \includegraphics[width= 1\textwidth]{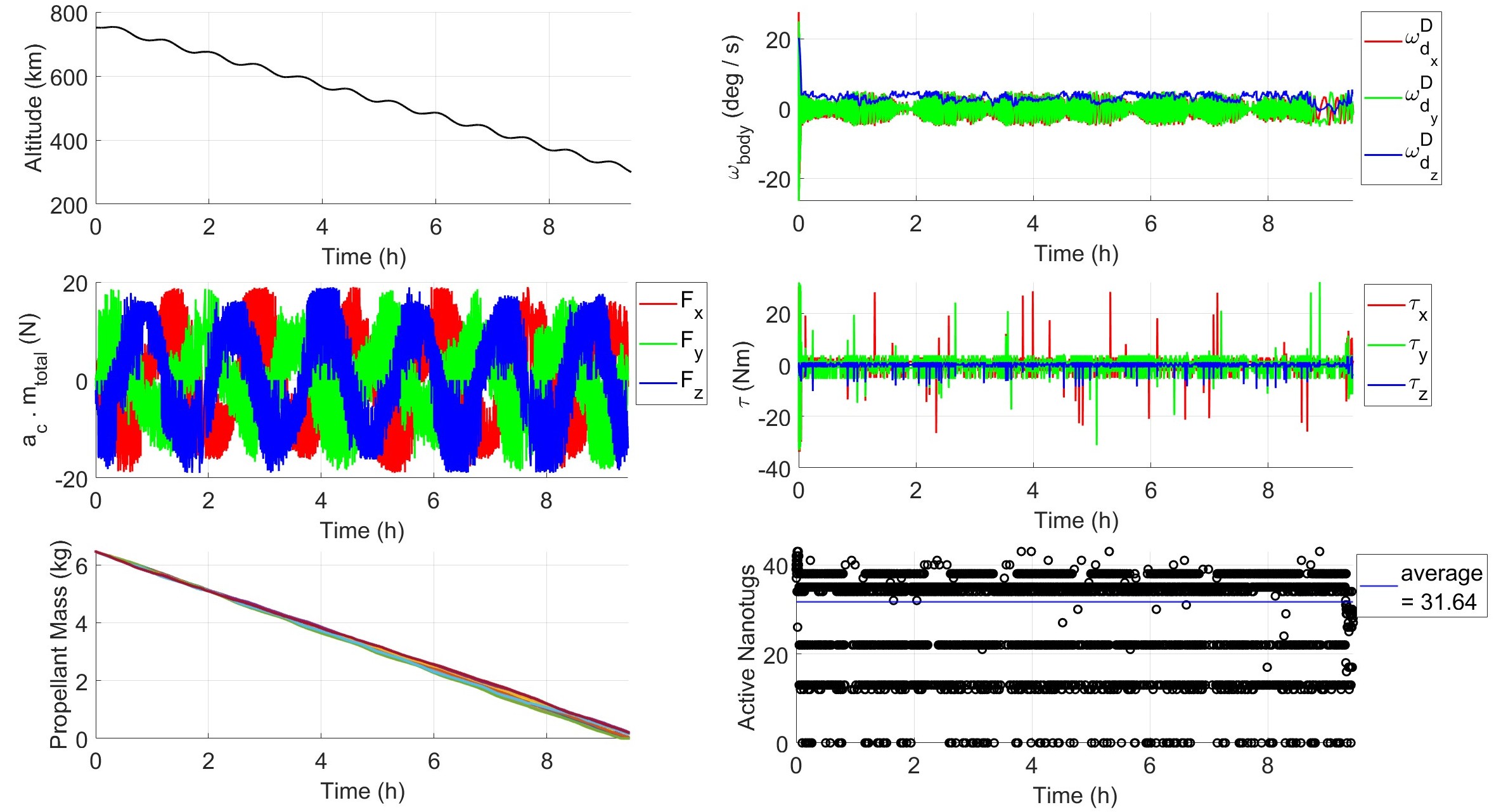}
    \caption{Simulation Results: Random Distribution.}
    \label{fig:randomResults}
\end{figure}

Figure~\ref{fig:randomResults} shows the results of the numerical simulation for the case of a random NanoTugs distribution. As can be seen from the figure, with the given parameters, the system was able to stabilize the debris quickly and subsequently de-orbit it within approximately 9–10~hours, while almost all of the propellant onboard the NanoTugs was consumed. The average number of NanoTugs operating simultaneously is $31.64$, which is consistent with the selected $\lambda$. Overall, the results are in good agreement with the analytical predictions in terms of de-orbiting time, propellant usage, and the number of active NanoTugs.

\begin{table}[h!]
\centering
\caption{Simulation Parameters: Random Distribution.}
\label{tab:simParametersRandom}
\begin{tabular}{l l l}
\Xhline{1pt}
\textbf{Parameter} & \textbf{Value} & \textbf{Units} \\
\hline
$\bm\omega_{d_0}^D$ & $[20, 20, 20]$ & deg/s \\

$N$ & $39 \times (1 + 10\%) = \lceil 42.9 \rceil = 43$ & - \\

$M_{i_\mathrm{prop}}$ & $5.8736 \times (1 + 10\%) = 6.46096$ & kg \\

$\bm\omega_r$ & $[0, 0, 0]$ & deg/s \\

$\delta\omega_\mathrm{max}$ & $5$ & deg/s \\

$K$ & $0$ & - \\

$\bm P$ & $\mathrm{diag}(500, 500, 500)$ & - \\
\Xhline{1pt}
\end{tabular}
\end{table}

\noindent\textit{Predefined NanoTugs Distribution} \\
The second scenario considers a predefined distribution of the NanoTugs with a targeted system attitude during de-orbiting. With this targeted attitude, it is assumed that all NanoTugs are simultaneously active for de-orbiting, and therefore $\lambda = 1$. As shown in Table~\ref{tab:systemSizing}, for this scenario and the given mission parameters, the required number of NanoTugs is 30, with each carrying $7.3420$~kg of propellant.

The numerical simulations are performed using the parameters listed in Table~\ref{tab:simParameterscont}. Similarly to the previous case, the initial angular velocity of the debris is set to $20$~deg/s along each axis. A $10\%$ margin is also applied to both the number of NanoTugs and the propellant mass per NanoTug. In this case, a nonzero control gain $K$ is introduced, along with higher gains on the angular velocity error compared to the random distribution scenario, to ensure that the desired attitude is maintained. The control torque is computed at each control step, as maintaining the targeted attitude is a primary objective in this scenario. The required attitude aligns $\bm e_x^D$ with the velocity direction, allowing all NanoTugs, given the chosen distribution, to provide thrust along $-\bm e_x^D$. The axis $\bm e_y^D$ is aligned with the orbit normal, and $\bm e_z^D = \bm e_x^D \times \bm e_y^D$, which is approximately in the radial direction for a near-circular orbit. This target attitude also minimizes the gravity gradient torque on the system, with $\bm e_z^D$ aligned along the debris longitudinal axis.

\begin{figure}[!h]
    \centering
    \includegraphics[width= 1\textwidth]{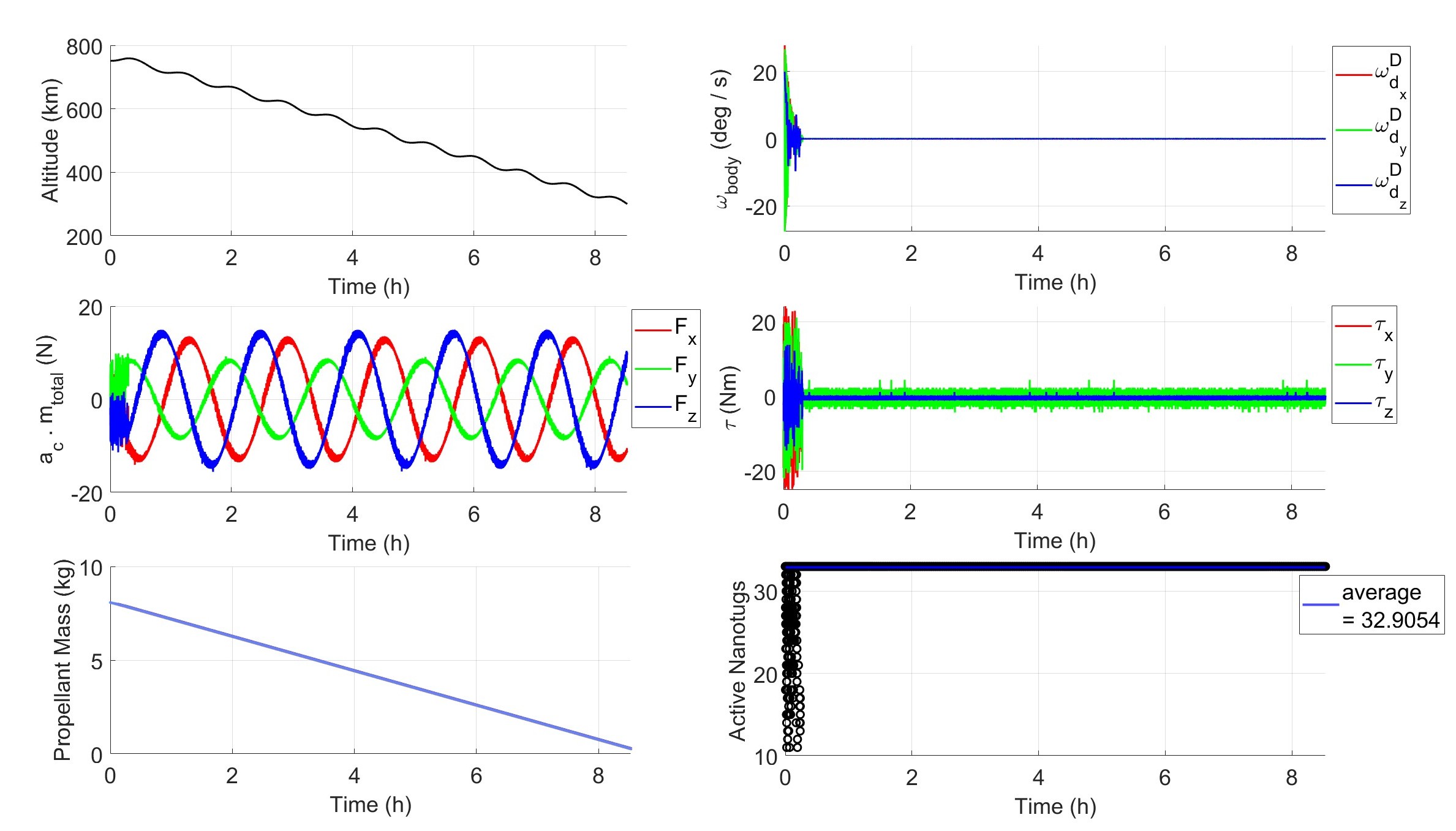}
    \caption{Simulation Results: Predefined Distribution.}
    \label{fig:controlledResults}
\end{figure}

Figure~\ref{fig:controlledResults} shows the results for this case. The de-orbiting is achieved within approximately 8–9~hours after detumbling, which takes less than half an hour. Approximately 9~kg of total propellant remains at the end of the mission. As observed in the figure, after detumbling, all NanoTugs are utilized at each time step. However, it should be noted that some NanoTugs are dedicated to attitude control to maintain the required pointing direction. Since it is assumed that each NanoTug can operate at most one thruster at a time, those used for attitude control cannot simultaneously use their thrusters aligned with the anti-velocity direction. This explains the need for margins in the analytical system sizing, which should be considered as an initial estimate rather than a precise prediction.\\

\begin{table}[h!]
\centering
\caption{Simulation Parameters: Predefined Distribution.}
\label{tab:simParameterscont}
\begin{tabular}{l l l}
\Xhline{1pt}
\textbf{Parameter} & \textbf{Value} & \textbf{Units} \\
\hline
$\bm\omega_{d_0}^D$ & $[20, 20, 20]$ & deg/s \\

$N$ & $30 \times (1 + 10\%) = 33$ & - \\

$M_{i_\mathrm{prop}}$ & $7.3420 \times (1 + 10\%) = 8.0762$ & kg \\

$\bm \omega_r$ & $[0, 0, 0]$ & deg/s \\

$K$ & $800$ & - \\

$\bm P$ & $\mathrm{diag}(1000, 1000, 1000)$ & - \\
\Xhline{1pt}
\end{tabular}
\end{table}

\noindent\textit{Results and Discussion} \\
As shown in the results, for both distributions, the selected and calculated system parameters were able to perform the de-orbiting mission within the expected time, with a margin of +1–2 hours. Each distribution strategy has certain advantages and disadvantages. Achieving a predefined distribution on a tumbling debris object can be challenging in practice, as it requires capturing specific surface points with high precision, which may not always be feasible. In contrast, a random distribution is easier to implement because it does not require precise targeting, making it more practical for operations where debris motion and properties are uncertain. Although the random distribution requires a higher total number of NanoTugs, the propellant per NanoTug is lower compared to the predefined distribution. In terms of control performance, the predefined distribution yields a more stable and predictable control behavior. In contrast, the random distribution results in oscillatory dynamics due to the reduced control authority, which introduces a mismatch between the control torque generated by the control law and the torque achievable by the swarm of NanoTugs. Moreover, the control authority can be limited given the thrusters operational constraints such as thruster response time and NanoTug power balance. In summary, a predefined distribution offers better control performance, whereas a random distribution is easier to implement in terms of debris capture operations. As a further refinement, one may consider that once one or a few NanoTugs have attached, a de-tumbling strategy could be applied to reduce the debris angular velocity, making the attachment of the remaining NanoTugs easier to achieve a predefined distribution.

This analysis presents cases where all the propellant on each NanoTug in the system is expected to be consumed, based on the analytical sizing approach. The concept was to match the total propellant mass of the system with the expected time to de-orbit assuming that all the propellant on each NanoTugs will be consumed and an $n$ NanoTugs will keep thrusting during this expected de-orbiting time, the analysis give the propellant per Nanotugs and based on that the required number of Nanotugs. 

There is a trade-off between the number of NanoTugs, the propellant per tug, and the time to de-orbit, where all are connected to the selected system parameters $I_{sp}$ and $T_i$. Thrusters with higher $I_{sp}$ generally provide lower thrust, which results in a longer mission duration but more efficient propellant use. On the other hand, selecting a thruster with lower $I_{sp}$ and higher thrust will perform de-orbiting faster but will require more propellant for the same mission. This is also connected to the control logic and the distribution of the NanoTugs. For example, in the discussed scenario, the required number of NanoTugs can be reduced by half while doubling the propellant mass of each unit. Using Eq.~\ref{eq: descent time, system sizing} with $M_{i_\mathrm{prop}} = 2M_{i_\mathrm{prop}}$, $n = n/2$, and $N = N/2$, this configuration leads to an approximately doubled de-orbiting time. This is because the total applied thrust to the system, $T_S = nT_i$, is reduced, resulting in a lower acceleration despite the increased available propellant per NanoTug.

On the other hand, studying the feasibility of performing multiple de-orbiting missions with the NanoTugs, as well as the feasibility of detaching from the debris and re-docking with the mother spacecraft, requires further analysis. In any case, these operations will require additional propellant per NanoTug, at least for re-docking with the mother spacecraft. Furthermore, the NanoTugs will also need to use their thrusters during the swarm establishment phase, which will require additional propellant.
\section{Conclusion} 
\label{sec:Conclusion}
This study explores an active debris removal mission concept using a swarm of nanosatellites, called NanoTugs. The NanoTugs are deployed by a mother spacecraft and attach to multiple points on the debris surface using a gripping mechanism. The primary objective of the paper was to evaluate the feasibility of a swarm-based approach for debris removal and to understand the relationship between swarm system size, mission parameters, and overall mission performance.

Control strategies and task allocation were incorporated for the stabilization and de-orbiting phases, showing that, with the proposed task allocation logic and the two NanoTugs distribution strategies, a tumbling debris object can be successfully de-orbited within required timeline and with expected propellant need while accounting for disturbance torques. The simulation examples also verified the analytical approach used for system sizing.

Future work can extend this study to additional mission phases and further improve the stabilization and de-orbiting performance. Potential improvements include developing an optimized task-allocation framework, enhanced cooperative control strategies, alternative NanoTugs distribution approaches, and power analysis considering shadowing effects from the debris.  Monte-Carlo analysis of debris de-orbiting scenarios simulation can be performed to assess the influence of the NanoTugs gripping location misalignment on the de-orbiting performance. Additionally, for large debris objects that may not fully burn up in the atmosphere, the impact location on Earth should be carefully planned, which could be addressed in future studies.

\end{document}